\documentclass[useAMS,usenatbib,usegraphicx]{mn2e}

\usepackage{longtable}

\newcommand{\grb}{\mbox{GRB\,130427A~}}
\newcommand{\grbnos}{\mbox{GRB\,130427A}}

\title[Comprehensive Radio View of \grbnos]{A Comprehensive Radio View of the Extremely Bright Gamma-Ray Burst 130427A}

\author[van der Horst et al.]{A.~J.~van~der~Horst,$^1$\thanks{e-mail: A.J.vanderHorst@uva.nl}, Z. Paragi$^{2}$, A.G. de Bruyn$^{3,4}$, J. Granot$^{5}$, C. Kouveliotou$^{6}$, \newauthor K. Wiersema$^{7}$, R.L.C. Starling$^{7}$, P.A. Curran$^{8}$, R.A.M.J. Wijers$^{1}$, A. Rowlinson$^{1}$, \newauthor G.A. Anderson$^{9,10}$, R.P. Fender$^{9,10}$, J. Yang$^{2,11}$, R.G. Strom$^{3}$\\
$^1$Anton Pannekoek Institute, University of Amsterdam, Science Park 904, 1098 XH Amsterdam, The Netherlands\\
$^{2}$Joint Institute for VLBI in Europe, Postbus 2, 7990 AA Dwingeloo, The Netherlands\\
$^{3}$ASTRON, the Netherlands Institute for Radio Astronomy, Postbus 2, 7990 AA Dwingeloo, The Netherlands\\
$^{4}$Kapteyn Astronomical Institute, PO Box 800, 9700 AV Groningen, The Netherlands\\
$^{5}$Departement of Natural Sciences, The Open University of Israel, P.O. Box 808, Ra'anana 43537, Israel\\
$^{6}$Space Science Office, ZP12, NASA/Marshall Space Flight Center, Huntsville, AL 35812, USA\\
$^{7}$Department of Physics and Astronomy, University of Leicester, University Road, Leicester LE1 7RH, UK\\
$^{8}$International Centre for Radio Astronomy Research $-$ Curtin University, GPO Box U1987, Perth, WA 6845, Australia\\
$^{9}$Department of Physics, Astrophysics, University of Oxford, Denys Wilkinson Building, Oxford, OX1 3RH, UK\\
$^{10}$School of Physics \& Astronomy, University of Southampton, Southampton, SO17 1BJ, UK\\
$^{11}$Department of Earth and Space Sciences, Chalmers University of Technology, Onsala Space Observatory, SE-43992 Onsala, Sweden}

\begin{document}

\maketitle

\label{firstpage}

\begin{abstract} 
\grb was extremely bright as a result of occurring at low redshift whilst the energetics were more typical of high-redshift gamma-ray bursts (GRBs).
We collected well-sampled light curves at 1.4 and 4.8~GHz of \grb with the Westerbork Synthesis Radio Telescope (WSRT); and we obtained its most accurate position with the European Very Long Baseline Interferometry Network (EVN). 
Our flux density measurements are combined with all the data available at radio, optical and X-ray frequencies to perform broadband modeling in the framework of a reverse-forward shock model and a two-component jet model, and we discuss the implications and limitations of both models. 
The low density inferred from the modeling implies that the \grb progenitor is either a very low-metallicity Wolf-Rayet star, or a rapidly rotating, low-metallicity O star. 
We also find that the fraction of the energy in electrons is evolving over time, and that the fraction of electrons participating in a relativistic power-law energy distribution is less than $15\%$. 
We observed intraday variability during the earliest WSRT observations, and the source sizes inferred from our modeling are consistent with this variability being due to interstellar scintillation effects. 
Finally, we present and discuss our limits on the linear and circular polarization, which are among the deepest limits of GRB radio polarization to date.
\end{abstract}


\section{Introduction} \label{section:intro}

Gamma-ray bursts (GRBs) are a broadband phenomenon, covering many orders of magnitude in observing frequency, from radio frequencies below 1~GHz to gamma-ray energies of tens of GeV. 
They also cover many orders of magnitude in observed timescales, from millisecond variability in the gamma-ray light curves up to months or even years at radio frequencies. 
Much of our understanding of the physics behind GRBs is based on multi-frequency and multi-timescale observations. 
In the case of long-duration GRBs \citep[i.e, with a duration $>2$~seconds;][]{kouveliotou1993} a picture has emerged in which a relativistic collimated outflow, or jet, is produced by a central engine, due to the collapse of a massive star \citep{woosley1993}; for short-duration GRBs most likely due to a binary merger of two compact objects \citep{eichler1989,narayan1992}. 
The prompt gamma-ray emission at keV to MeV energies is believed to be produced by particles accelerated in shocks internal to the outflow, while the later time afterglow emission (from X-ray to radio frequencies, and arguably also the long-lasting GeV gamma-ray emission), is due to the interaction of the jet with the ambient medium \citep[see][for recent reviews]{kouveliotou2012}. 
At the front of the jet, matter is swept up and a forward shock is formed, accompanied by a short-lived reverse shock moving back into the outflow. 
The forward shock is initially moving at relativistic speeds but decelerating, while the reverse shock can be either relativistic or Newtonian. 
The observed afterglows are usually dominated by emission from the forward shock, but occasionally the reverse shock causes a bright optical flash peaking in the first minutes and a radio flare in the first days after the GRB onset \citep[e.g.][]{akerlof1999,kulkarni1999}. 
Radio observations are important for constraining the spectra and evolution of the forward and reverse shocks, and follow the evolution of the GRB jet up to much later times than at higher frequencies \citep[for a recent review on GRB radio observations and their implications for GRB jet physics, see][]{granot2014}. 

Over the last decade new ground- and space-based observatories have provided broadband GRB data sets, e.g. the {\it Fermi Gamma-ray Space Telescope} for detecting high-energy gamma-rays, the {\it Swift} satellite for X-ray light curves, robotic optical telescopes for early-time light curves, and improved and new facilities for observations at radio frequencies. 
However, it is quite rare that excellent broadband coverage is accompanied with great temporal sampling, in particular at the extreme ends of the spectrum \citep[e.g.][]{cenko2011}; conversely, some GRBs with extremely well sampled light curves do not have comparable spectral coverage \citep[e.g.][]{racusin2008}. 
The recent, extremely bright, long-duration \grb was the exception that brought all these observational capabilities together, from its detection in gamma-rays to its multi-wavelength follow-up observations.

Most long-duration GRBs occur at high redshifts, with a mean redshift at $z\simeq2$ \citep{fynbo2009,jakobsson2012}; the current record holder is at $z\simeq9.4$ \citep{cucchiara2011}. 
For a small group of these at low redshifts ($z<0.4$), we are able to detect and identify spectroscopically their associated supernovae, although this does not always appear to be the case \citep[e.g.,][]{fynbo2006,gehrels2006}. 
A significant fraction of that group has intrinsic luminosities and energetics lower than those of GRBs at higher redshifts \citep[e.g.][]{kaneko2007,starling2011}; even the most luminous one to date, GRB\,030329, is at the low end of the energetics distribution for the total GRB sample \citep{kaneko2007}. 
\grb is exceptional in that, although it is at a low redshift of $z=0.34$, with an accompanying supernova of the same type as the other GRB-associated supernovae \citep[SN\,2013cq;][]{levan2013,xu2013}, it is comparable in luminosity to the majority of long GRBs. 
At gamma-ray energies this is a record-breaking GRB, with the highest observed fluence in 29 years, the longest lasting high-energy gamma-ray afterglow (i.e. 20~hours), and the highest energy gamma-ray photon ever detected \citep[95~GeV;][]{ackermann2014}. 
Compared to the entire GRB sample, the \grb X-ray and optical observed brightness are amongst the highest, while its intrinsic luminosities are just above or around the average \citep{perley2014}. 
Given the extremely well sampled light curves for \grbnos, and the fact that the light curves at X-ray and optical frequencies are comparable to those of other high-luminosity GRBs, this source provides a unique opportunity to study not only the physics of this particular GRB in great detail \citep[e.g.,][]{kouveliotou2013,preece2014}, but also to make inferences for GRBs at more typical redshifts.

A remarkable feature of \grb is the early-time peak at optical frequencies, $\sim10-20$~seconds after the GRB onset, for which an optical flash due to the reverse shock has been suggested as the most likely explanation \citep{vestrand2014}. 
At radio frequencies the light curves display a peak on a day timescale, which has also been attributed to the reverse shock \citep[][]{laskar2013,perley2014,anderson2014}. 
Broadband modeling efforts have shown that the light curves from radio to X-ray frequencies, and also the high-energy gamma-ray light curves, can indeed be interpreted as a combination of emission from the forward and reverse shocks \citep{laskar2013,panaitescu2013,maselli2014,perley2014}. 
In this paper we present radio observations of \grb with the Westerbork Synthesis Radio Telescope (WSRT) at two radio frequencies (Section~\ref{sec:wsrt}), resulting in well sampled light curves and enabling more detailed modeling than previous efforts. 
We also show the results from Very Long Baseline Interferometry (VLBI) observations with the European VLBI Network (EVN), which set constraints on the source size and provide the best localization of this GRB  (Section~\ref{sec:vlbi}). 
We revisit the modeling of the broadband light curves to set more stringent constraints on the evolution of the forward and reverse shock spectra, and present a two-component jet model as an alternative to describe all the available data from radio to X-ray frequencies (Section~\ref{sec:model}). 
Since our WSRT observations have long durations, we also present radio brightness variations at relatively short timescales to study variability of the source and possible scintillation effects (Section~\ref{sec:shvar}). 
Furthermore, due to the source brightness we can put very tight constraints on the linear and circular radio polarization, and discuss those in the context of GRB afterglow emission models (Section~\ref{sec:pola}). 
Finally, we summarize our results and draw some conclusions (Section~\ref{sec:concl}).

\section{WSRT Observations}\label{sec:wsrt}

\begin{figure}
\begin{center}
\includegraphics[angle=-90,width=\columnwidth]{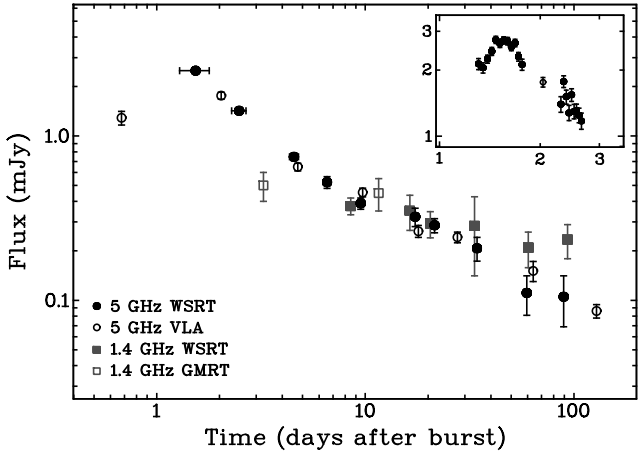}
\caption{Radio light curves at 1.4~GHz (squares) and 4.8~GHz (circles) of \grbnos. 
The solid symbols are the WSRT measurements presented in this paper, 
while the open symbols are the VLA and GMRT results from \citet{laskar2013} and \citet{perley2014}. 
The inset shows the detailed flux density evolution at 4.8~GHz during the first 3 days, using WSRT images made with 1-hour integration times.}
\label{fig:radiolcs}
\end{center}
\end{figure}

\begin{table}
\begin{center}
\caption{WSRT observations of \grbnos, with $\Delta$T the midpoint of each observation in days after the {\it Fermi}/GBM trigger time. 
The long 4.8~GHz observations on April 28/29 and 29/30 have been divided up into 1-hour time intervals and the results are given at the bottom of the table. 
\label{tab:radio}}
\renewcommand{\arraystretch}{1.1}
\begin{tabular}{|l|c|c|c|c|} 
\hline
Epoch & $\Delta$T & Int. time & Freq. & Flux \\
 & (days) & (hours) & (GHz) & ($\mu$Jy) \\
\hline
Apr 28.611 $-$ 29.110 & 1.52 & 12.0 & 4.8 & 2500$\pm$25 \\
Apr 29.608 $-$ 30.001 & 2.47 & 9.4 & 4.8 & 1424$\pm$24 \\
May 1.651 $-$ 2.102$^{\,\rm{a}}$ & 4.55 & 5.4 & 1.4 & 283$\pm$711 \\
May 1.651 $-$ 2.102 & 4.55 & 5.4 & 4.8 & 746$\pm$37 \\
May 3.660 $-$ 4.097$^{\,\rm{b}}$ & 6.55 & 10.5 & 4.8 & 523$\pm$43 \\
May 5.592 $-$ 6.091 & 8.51 & 12.0 & 1.4 & 375$\pm$44 \\
May 6.592 $-$ 7.088 & 9.51 & 12.0 & 4.8 & 389$\pm$31 \\
May 13.570 $-$ 13.796 & 16.36 & 5.4 & 1.4 & 351$\pm$85 \\
May 14.567 $-$ 14.793 & 17.36 & 5.4 & 4.8 & 322$\pm$41 \\
May 17.559 $-$ 18.058 & 20.48 & 12.0 & 1.4 & 293$\pm$53 \\
May 18.557 $-$ 19.055 & 21.48 & 12.0 & 4.8 & 286$\pm$28 \\
May 30.524 $-$ 30.855 & 33.36 & 8.0 & 1.4 & 284$\pm$143 \\
May 31.521 $-$ 31.852 & 34.36 & 8.0 & 4.8 & 207$\pm$34 \\
Jun 25.453 $-$ 25.951 & 59.38 & 12.0 & 4.8 & 111$\pm$30 \\
Jun 26.450 $-$ 26.949 & 60.37 & 12.0 & 1.4 & 209$\pm$51 \\
Jul 25.371 $-$ 25.869 & 89.30 & 12.0 & 4.8 & 105$\pm$36 \\
Jul 29.360 $-$ 29.858 & 93.28 & 12.0 & 1.4 & 234$\pm$55 \\
\hline
Apr 28.611 $-$ 28.653 & 1.31 & 1.0 & 4.8 & 2132$\pm$124 \\
Apr 28.653 $-$ 28.694 & 1.35 & 1.0 & 4.8 & 2047$\pm$108 \\
Apr 28.694 $-$ 28.736 & 1.39 & 1.0 & 4.8 & 2244$\pm$95 \\
Apr 28.736 $-$ 28.777 & 1.43 & 1.0 & 4.8 & 2433$\pm$98 \\
Apr 28.777 $-$ 28.819 & 1.47 & 1.0 & 4.8 & 2743$\pm$101 \\
Apr 28.819 $-$ 28.860 & 1.51 & 1.0 & 4.8 & 2640$\pm$105 \\
Apr 28.860 $-$ 28.902 & 1.56 & 1.0 & 4.8 & 2728$\pm$101 \\
Apr 28.902 $-$ 28.943 & 1.60 & 1.0 & 4.8 & 2707$\pm$107 \\
Apr 28.943 $-$ 28.985 & 1.64 & 1.0 & 4.8 & 2551$\pm$103 \\
Apr 28.985 $-$ 29.026 & 1.68 & 1.0 & 4.8 & 2654$\pm$105 \\
Apr 29.026 $-$ 29.068 & 1.72 & 1.0 & 4.8 & 2300$\pm$102 \\
Apr 29.068 $-$ 29.110 & 1.76 & 1.0 & 4.8 & 2117$\pm$121 \\
\hline
Apr 29.608 $-$ 29.652 & 2.31 & 1.0 & 4.8 & 1399$\pm$113 \\
Apr 29.652 $-$ 29.695 & 2.35 & 1.0 & 4.8 & 1773$\pm$110 \\
Apr 29.695 $-$ 29.739 & 2.39 & 1.0 & 4.8 & 1511$\pm$111 \\
Apr 29.739 $-$ 29.782 & 2.43 & 1.0 & 4.8 & 1278$\pm$101 \\
Apr 29.782 $-$ 29.826 & 2.48 & 1.0 & 4.8 & 1543$\pm$102 \\
Apr 29.826 $-$ 29.869 & 2.52 & 1.0 & 4.8 & 1298$\pm$99 \\
Apr 29.869 $-$ 29.913 & 2.57 & 1.0 & 4.8 & 1311$\pm$87 \\
Apr 29.913 $-$ 29.956 & 2.61 & 1.0 & 4.8 & 1247$\pm$95 \\
Apr 29.956 $-$ 30.001 & 2.65 & 1.0 & 4.8 & 1174$\pm$99 \\
\hline
\end{tabular}
\begin{list}{}{}
\item[$^{\rm a}$] Non-detection, not shown in Figure~\ref{fig:radiolcs} \
\item[$^{\rm b}$] Part of the EVN run \\
\end{list}
\end{center}
\end{table}

We observed \grb at 1.4 and 4.8~GHz with the WSRT from 28 April to 29 July 2013. 
We used the Multi Frequency Front Ends \citep{tan1991} in combination with the IVC+DZB back end in continuum mode, with a bandwidth of 8x20 MHz at both observing frequencies. 
Gain and phase calibrations were performed with the calibrator 3C~286 for all observations. 
The observations were analyzed using the Multichannel Image Reconstruction Image Analysis and Display 
\citep[MIRIAD;][]{sault1995} software package. 
The observing dates, integration times, and flux density measurements of our observations are listed in Table~\ref{tab:radio}. 
Figure~\ref{fig:radiolcs} shows the light curves at our observing frequencies together with the VLA and GMRT flux densities at the same frequencies \citep{laskar2013,perley2014}. 

Since WSRT is an East-West array, with all the dishes placed along one line so that the Earth's rotation is used to fill the {\it uv} plane, it is common to observe for several (up to 12) hours for making high quality images. 
Given the brightness of \grb in the first two epochs at 4.8~GHz we were able to make multiple images by dividing the long observations into shorter time intervals, i.e. of 1~hour duration, after subtracting all the other sources in the field using the MIRIAD task \texttt{uvmodel}. 
The resulting flux densities are reported at the lower half of Table~\ref{tab:radio} and shown in the inset of Figure~\ref{fig:radiolcs}. 
We also fit a point source to the visibility data with the MIRIAD task \texttt{uvfit} at 15-minute intervals after subtracting the other sources in the field. 
The flux densities we obtained in these two different ways will be discussed in Section~\ref{sec:shvar}.

The exceptional brightness of \grb during the first few days allowed linear and circular polarization searches. 
We made images in Stokes~Q, U, and V, but we did not detect significant emission at the position of the GRB. 
The formal flux density measurements and $3\sigma$ upper limits for the first three epochs are given in Table~\ref{tab:pola}. 
We combined these with the Stokes~I values reported in Table~\ref{tab:radio} and determined upper limits on the linear polarization $P_{\rm{L}}$ and circular polarization $P_{\rm{C}}$. 
Table~\ref{tab:pola} shows that these limits are only a few to several percent at the first two epochs, with the most stringent limits being $P_{\rm{L}}<3.9\%$ and  $P_{\rm{C}}<2.7\%$ in the first epoch; in the third epoch the polarization limits are more than $10\%$. 
As the source becomes significantly fainter at later times, the polarization limits get higher (tens of percent) and not constraining for emission models, so these are therefore not reported here.

\begin{table*}
\begin{center}
\caption{Polarization limits on \grb for the first three epochs at 4.8~GHz: the $3\sigma$ upper limits and formal flux density measurements (between parentheses) for a point source at the position of the GRB in the Stokes Q, U and V images; and the resulting limits on the linear polarization $P_{\rm{L}}$ and circular polarization $P_{\rm{C}}$.}
\label{tab:pola}
\renewcommand{\arraystretch}{1.1}
\begin{tabular}{|l|c|c|c|c|c|} 
\hline
Epoch & Q & U & V & $P_{\rm{L}}$ & $P_{\rm{C}}$ \\
 & ($\mu$Jy) & ($\mu$Jy) & ($\mu$Jy) & (\%) & (\%) \\
\hline
Apr 28.611 $-$ 29.110 & $<66$ ($8\pm22$) & $<66$ ($57\pm22$) & $<66$ ($62\pm22$) & $<3.9$ & $<2.7$ \\
Apr 29.608 $-$ 30.001 & $<69$ ($0\pm23$) & $<72$ ($16\pm24$) & $<75$ ($22\pm25$) & $<7.5$ & $<5.7$ \\
May 1.651 $-$ 2.102 & $<90$ ($6\pm30$) & $<87$ ($12\pm29$) & $<90$ ($2\pm30$) & $<21$ & $<15$ \\
\hline
\end{tabular}
\end{center}
\end{table*}

\section{EVN Observations}\label{sec:vlbi}

\grb was observed with the EVN at 5~GHz from 15:50~UT on 3 May 2013 until 02:20~UT on 4 May 2013. 
Participating telescopes were Arecibo, Effelsberg, Jodrell Bank (MkII), Medicina, Noto, Onsala, Sheshan, Torun, Yebes and WSRT (see Table~\ref{tab:EVNtels} for telescope parameters). 
The 2-bit sampled data were streamed from most telescopes to the EVN Software Correlator at JIVE (SFXC) at a rate of 1024~Mbit/s/telescope. 
Arecibo and Shanghai sent 1-bit sampled data at a rate of 512~Mbit/s. 
The nearby compact calibrator J1134+2901 was used as phase-reference during the observations. 
The telescopes were switching rapidly between the phase-reference and the target, separated by 1.4 degrees, in 1:30--3:30 minutes cycles. 
The data were calibrated using standard procedures in the Astronomical Image Processing System \citep[AIPS, e.g.,][]{vanmoorsel1996}. 

\grb was detected with a peak brightness of 460~$\mu$Jy/beam at the position of RA=11$^{\rm h}$ 32$^{\rm m}$ 32.80872$^{\rm s}$, Dec=+27\degr 41' 56.0203" (J2000), with an estimated error of 0.6~mas. 
The naturally weighted restoring beam was 3.4$\times$0.9~milliarcsecond (mas), with major axis position angle $-49\degr$. 
Fitting a circular Gaussian model to the $uv$-data in Difmap \citep{shepherd1994} resulted in a source size of 0.6~mas and a total flux density of 550~$\mu$Jy. 
A point source fit to the VLBI data resulted in 460$\pm$50~$\mu$Jy total flux density as measured by the EVN, consistent with the flux density measured by the WSRT independently. 
The errors include statistical (rms noise 18~$\mu$Jy/beam) and systematic components ($\sim10\%$ amplitude calibration accuracy). 

We consider 0.6~mas to be an upper limit on the source size, because of the residual phase and amplitude errors that might still be present in the data. 
We did also observe two very nearby radio sources as candidate secondary calibrators. 
One of these was not detected above the 5$\sigma$ noise level, the other was detected only at the $\sim10\sigma$ level. 
Therefore, we could not further improve on the phase calibration. 
At the redshift $z=0.34$ of \grb an angular size of 1~mas corresponds to a physical size of $1.49\times10^{19}$~cm, which means that the upper limit on the source size from our EVN observation is $9\times10^{18}$~cm at 6.55~days. 
For a circular expanding source this corresponds to an average expansion speed of $<265$~c, which is not very constraining several days after the GRB onset, since by that time the Lorentz factor is typically a few tens at most (see also Section~\ref{sec:modelrevfor}).

\begin{table}
\begin{center}
\caption{Parameters of the telescopes participating in the EVN observations}
\label{tab:EVNtels}
\renewcommand{\arraystretch}{1.1}
\begin{tabular}{|l|c|c|}
\hline
Radio telescope    & Diameter (m) & SEFD$^{\,\rm{a}}$ (Jy)  \\
\hline
Arecibo     & 305 & 5 \\
Effelsberg & 100 & 20 \\
Jodrell Bank MkII & 25 & 320 \\
Medicina   & 32   & 170 \\
Noto           & 32   & 260 \\
Onsala       & 25   & 600 \\
Sheshan (Shanghai) & 25 & 720 \\
Torun       & 32   & 220 \\
Yebes        & 40   & 160 \\
WSRT        & $12\times 25^{\,\rm{b}}$ & 120 \\
\hline
\end{tabular}
\begin{list}{}{}
\item[$^{\rm a}$] System Equivalent Flux Density \
\item[$^{\rm b}$] The telescope was used in phased array mode for the VLBI \
\item[\,\,\,] observations, but also produced local interferometer data \\
\end{list}
\end{center} 
\end{table}

\section{Modeling}\label{sec:model}

The wealth of data on \grb accumulated across the electromagnetic spectrum has enabled a detailed broadband modeling beyond what has been done before for any GRB. 
Here we build on the modeling results that have already been presented in the literature \citep{kouveliotou2013,laskar2013,panaitescu2013,maselli2014,bernardini2014,perley2014}, by not only adding the radio observations presented in the previous section, and discussing their implications, but also by examining the various assumptions in, and inferences from, previous modeling efforts. 
For this purpose we have combined our WSRT results with all the radio, optical and X-ray data available in the literature \citep{laskar2013,anderson2014,maselli2014,perley2014,vestrand2014}. 
We did not include the high-energy gamma-ray data from the {\it Fermi}/LAT, although we did use some inferences made from the optical to gamma-ray spectra \citep{kouveliotou2013}.

\subsection{Broadband Spectra}\label{sec:spec}

We discuss here the implications of the broadband spectra, without considering information from the light curves. 
GRB afterglow spectra are usually described in terms of broadband synchrotron emission produced by electrons which are accelerated by a strong shock. 
These spectra are characterized by four power-law segments with three break frequencies \citep{sari1998}: the peak frequency $\nu_{\rm{m}}$, the cooling frequency $\nu_{\rm{c}}$, and the synchrotron self-absorption frequency $\nu_{\rm{a}}$. 
These three frequencies can be ordered in various ways, but the most relevant for this discussion are $\nu_{\rm{a}}<\nu_{\rm{m}}<\nu_{\rm{c}}$ and $\nu_{\rm{m}}<\nu_{\rm{a}}<\nu_{\rm{c}}$. 
In the former case the spectral power-law index in between $\nu_{\rm{a}}$ and $\nu_{\rm{m}}$ is $\beta=1/3$ (with the flux $F_{\nu}\propto\nu^{\beta}$), and in the latter case $\beta=5/2$ in between $\nu_{\rm{m}}$ and $\nu_{\rm{a}}$. 
In both cases $\beta=2$ below all three characteristic frequencies, $\beta=-(p-1)/2$ in between $\nu_{\rm{a,m}}$ and $\nu_{\rm{c}}$, and $\beta=-p/2$ above $\nu_{\rm{c}}$. 
The parameter $p$ is the power-law index of the energy distribution of the synchrotron emitting electrons. 
From these three characteristic frequencies and the peak flux $F_{\nu,\rm{max}}$ one can determine four physical parameters: the isotropic equivalent kinetic energy $E$ of the shock, the density $\rho$ of the medium that the shock is moving through, and the fractions $\varepsilon_{\rm{e}}$ and $\varepsilon_{\rm{B}}$ of the internal energy density in electrons and the magnetic field, respectively.

\citet{laskar2013} have compiled broadband spectra for \grb at various epochs, including radio, near-infrared, optical and X-ray data, and shown that these can not be explained by a single synchrotron spectrum as one would expect from a GRB blast wave. 
This has been confirmed by \citet{perley2014} for more epochs, by using more data, and also including high-energy gamma-ray observations. 
While the optical to gamma-ray spectra can be explained by a broken power law with typical slopes for GRB afterglows \citep[see also][]{kouveliotou2013}, the radio spectra are more complex: at most epochs they do not show any of the characteristic spectral slopes, but are in fact fairly flat, i.e., $\beta\simeq0$. 
Only at $0.6-0.7$~days there is a spectral turn-over at the low radio frequencies, with a steep spectral index $\beta\simeq2.4$ between 5.1 and 6.8~GHz \citep{laskar2013,perley2014}, and a less steep $\beta\simeq1$ between 5.1 and 15.7~GHz \citep{anderson2014}. 
The instantaneous broadband spectra at various epochs imply that there are two spectral components: one with the peak at $\nu_{\rm{m}}$, and another one at lower frequencies where self-absorption plays a significant role. 
The self-absorption frequency $\nu_{\rm{a}}$ of the high-frequency component can not be constrained since the second component is dominating the emission at low frequencies. 

The evolution of the near-infrared to optical spectra also suggests the presence of two components. 
\citet{perley2014} have shown that the optical spectral index evolves from $-0.3$ to $-0.4$ in the first day, to $-0.7$ after a few days. 
This latter spectral index is the same as the spectral index derived from spectral fits at 1.5 and 5~days including near-infrared to high-energy gamma-ray data \citep{kouveliotou2013}. 
The latter fits do require a spectral break with a slope change of 0.5, characteristic of the cooling break $\nu_{\rm{c}}$, at a few tens of keV. 
This $\nu_{\rm{c}}$ value is just above the {\it Swift} X-Ray Telescope (XRT) observing band \citep{kouveliotou2013}, and was measured largely using {\it NuSTAR} observations; spectral fits of the {\it Swift}/XRT data alone also resulted in $\beta=-0.7$ \citep{maselli2014}. 
The softer near-infrared to optical spectra at early times can be explained by a contribution from both aforementioned spectral components. 
To cause this particular evolution from a soft to a harder spectrum, the peak of the high-frequency spectral component should be initially above the optical regime and then move down through the observing bands, while the peak of the low-frequency component is initially already below the near-infrared frequencies. 
Once the peak of the high-frequency component has moved below the near-infrared frequencies as well, the spectrum becomes optically thin.

\subsection{Light Curves}\label{sec:lcs}

\begin{table}
\begin{center}
\caption{Temporal power-law indices of the radio (15~GHz), optical (R band), and X-ray light curves presented in Figures~\ref{fig:modelrevfor} and~\ref{fig:modeldouble}.}
\label{tab:lcdecay}
\renewcommand{\arraystretch}{1.1}
\begin{tabular}{|l|l|c|}
\hline
Frequency & Time range & Temporal \\
regime & (days) & index \\
\hline
Radio & $0.3-0.7$ & $0.33\pm0.20$ \\
 & $0.7-4$ & $-1.16\pm0.14$ \\
 & $4-60$ & $-0.48\pm0.07$ \\
\hline
Optical & $0.00007-0.0002$ & $1.44\pm0.08$ \\
 & $0.0002-0.001$ & $-1.87\pm0.08$ \\
 & $0.001-0.004$ & $-0.85\pm0.01$ \\
 & $0.004-0.02$ & $-1.20\pm0.01$ \\
 & $0.02-0.6$ & $-0.91\pm0.01$ \\
 & $0.6-40$ & $-1.33\pm0.01$ \\
\hline
X-rays & $0.005-0.02$ & $-1.30\pm0.01$ \\
 & $0.2-180$ & $-1.35\pm0.01$ \\
\hline
\end{tabular}
\end{center} 
\end{table}

The light curves at various observing frequencies are determined by the evolution of the characteristic frequencies and the peak flux. 
These are governed by the evolution and dynamics of the shocks that produce the synchrotron emission of both aforementioned spectral components. 
Modeling of \grb has been performed \citep{laskar2013,panaitescu2013,maselli2014,perley2014} by assuming that the high-frequency spectral component is the forward shock moving into the ambient medium, while the low-frequency component is the reverse shock moving back into the outflow. 
These modeling efforts, however, were not based on the full data set available now, in particular the well-sampled radio light curves presented in this paper and \citet{anderson2014}. 
We will first discuss the reverse-forward shock model as proposed by other authors, the assumptions that have been made, and how well it fits the broadband light curves. 
We will then present a two-component jet model as an alternative to fit these light curves. 
The latter model also requires reverse shock emission to explain the observed optical flash \citep{vestrand2014}, but the low-frequency spectral component is explained by emission similar to that of a forward shock. 
Both models require an extra ingredient to account for the very fast evolution of the peak of the spectrum from optical to radio frequencies, namely time-varying microphysical parameters. 

The best sampled radio light curves, at 1.4, 5, 7, 15, 36 and 90~GHz, are shown in Figures~\ref{fig:modelrevfor} and \ref{fig:modeldouble}, together with optical light curves in the I- and R-band, and the X-ray light curve at 3~keV. 
The R-band is the only near-infrared/optical/UV band with early enough coverage to show the initial rise of the light curve, followed by several phases of steep decay and flattening. 
Power-law indices for various segments of the R-band light curve are given in Table~\ref{tab:lcdecay}. 
The X-ray light curve shows the very steep decay typical of high-latitude prompt emission, with the afterglow emission dominating after 0.005~d. 
The observed X-ray light curve has similar decay slopes as the R-band light curve, which are also shown in Table~\ref{tab:lcdecay}, but power-law fits to the light curve sections before and after the gap between 0.02 and 0.2~d show that there is a different normalization \citep[and not a jet break as suggested by][]{maselli2014}, indicating that in this gap a flattening of the light curve also occurred at X-ray frequencies. 
At the other side of the spectrum, the radio light curves show a rise, in particular at 5 and 15~GHz, followed by a decay similar to the one observed at optical and X-ray frequencies, and also a flattening followed by a steeper decay (see Table~\ref{tab:lcdecay} for the temporal indices at 15~GHz). 
The power-law index of steeper decay can not be well constrained due to a lack of late-time observations with the required sensitivity.

\begin{table*}
\begin{center}
\caption{Temporal power-law indices of $F_{\nu,\rm{max}}$, $\nu_{\rm{m}}$ and $\nu_{\rm{c}}$, and $F_{\nu}$ in various spectral regimes, for relativistic forward shocks \citep{vanderhorst2007}, and thick-shell \citep[relativistic;][]{kobayashi2000,chevalier2000,yi2013} and thin-shell \citep[Newtonian;][]{kobayashi2000,zou2005} reverse shocks. The temporal power-law indices in this table depend on the power-law index $p$ of the electron energy distribution, the power-law index $k$ of the ambient medium density with radius, and the power-law index $g$ of the Lorentz factor as a function of radius for thin-shell reverse shocks.}
\label{tab:parscale}
\renewcommand{\arraystretch}{1.1}
\begin{tabular}{|l|c|c|c|}
\hline
 & Forward shock & \multicolumn{2}{c}{Reverse shock} \\
 & & Thick-shell & Thin-shell \\
\hline \vspace{4pt}
$F_{\nu,\rm{max}}$ & $-\frac{k}{2(4-k)}$ & $-\frac{47-10k}{12(4-k)}$ & $-\frac{11g+12}{7(2g+1)}$ \\ \vspace{4pt}
$\nu_{\rm{c}}$ & $-\frac{4-3k}{2(4-k)}$ & $-\frac{73-14k}{12(4-k)}$ & $-\frac{3(5g+8)}{7(2g+1)}$ \\ \vspace{4pt}
$\nu_{\rm{m}}$ & $-\frac{3}{2}$ & $-\frac{73-14k}{12(4-k)}$ & $-\frac{3(5g+8)}{7(2g+1)}$ \\ \vspace{4pt}
$\nu_{\rm{a}}$ ($\nu_{\rm{a}}<\nu_{\rm{c}}<\nu_{\rm{m}}$) & $-\frac{10+3k}{5(4-k)}$ & $-\frac{32-7k}{15(4-k)}$ & $-\frac{3(11g+12)}{35(2g+1)}$ \\ \vspace{4pt}
$\nu_{\rm{a}}$ ($\nu_{\rm{a}}<\nu_{\rm{m}}<\nu_{\rm{c}}$) & $-\frac{3k}{5(4-k)}$ & $-\frac{32-7k}{15(4-k)}$ & $-\frac{3(11g+12)}{35(2g+1)}$ \\ \vspace{2pt}
$\nu_{\rm{a}}$ ($\nu_{\rm{m}}<\nu_{\rm{a}}<\nu_{\rm{c}}$) & $-\frac{3p(4-k)+2(4+k)}{2(4-k)(p+4)}$ & $-\frac{p(73-14k)+2(67-14k)}{12(4-k)(p+4)}$ & $-\frac{3p(5g+8)+8(4g+5)}{7(2g+1)(p+4)}$ \\
\hline \vspace{4pt}
$F_{\nu}$ ($\nu<\nu_{\rm{a}}<\nu_{\rm{c}}<\nu_{\rm{m}}$) & $\frac{4}{4-k}$ & $\frac{5-k}{3(4-k)}$ & $\frac{5g+8}{7(2g+1)}$ \\ \vspace{4pt}
$F_{\nu}$ ($\nu_{\rm{a}}<\nu<\nu_{\rm{c}}<\nu_{\rm{m}}$) & $\frac{2-3k}{3(4-k)}$ & $-\frac{17-4k}{9(4-k)}$ & $-\frac{2(3g+2)}{7(2g+1)}$ \\ \vspace{4pt}
$F_{\nu}$ ($\nu_{\rm{a}}<\nu_{\rm{c}}<\nu<\nu_{\rm{m}}$) & $-\frac{1}{4}$ & $-\frac{167-34k}{24(4-k)}$ & $-\frac{37g+48}{14(2g+1)}$ \\ \vspace{2pt}
$F_{\nu}$ ($\nu_{\rm{a}}<\nu_{\rm{c}}<\nu_{\rm{m}}<\nu$) & $-\frac{3p-2}{4}$ & $-\frac{p(73-14k)+2(47-10k)}{24(4-k)}$ & $-\frac{3p(5g+8)+2(11g+12)}{14(2g+1)}$ \\
\hline \vspace{4pt}
$F_{\nu}$ ($\nu<\nu_{\rm{a}}<\nu_{\rm{m}}<\nu_{\rm{c}}$) & $\frac{2}{4-k}$ & $\frac{5-k}{3(4-k)}$ & $\frac{5g+8}{7(2g+1)}$ \\ \vspace{4pt}
$F_{\nu}$ ($\nu_{\rm{a}}<\nu<\nu_{\rm{m}}<\nu_{\rm{c}}$) & $\frac{2-k}{4-k}$ & $-\frac{17-4k}{9(4-k)}$ & $-\frac{2(3g+2)}{7(2g+1)}$ \\ \vspace{4pt}
$F_{\nu}$ ($\nu_{\rm{a}}<\nu_{\rm{m}}<\nu<\nu_{\rm{c}}$) & $-\frac{3p(4-k)-12+5k}{4(4-k)}$ & $-\frac{p(73-14k)+3(7-2k)}{24(4-k)}$ & $-\frac{3p(5g+8)+7g}{14(2g+1)}$ \\ \vspace{2pt}
$F_{\nu}$ ($\nu_{\rm{a}}<\nu_{\rm{m}}<\nu_{\rm{c}}<\nu$) & $-\frac{3p-2}{4}$ & $-\frac{p(73-14k)+2(47-10k)}{24(4-k)}$ & $-\frac{3p(5g+8)+2(11g+12)}{14(2g+1)}$ \\
\hline \vspace{4pt}
$F_{\nu}$ ($\nu<\nu_{\rm{m}}<\nu_{\rm{a}}<\nu_{\rm{c}}$) & $\frac{2}{4-k}$ & $\frac{5-k}{3(4-k)}$ & $\frac{5g+8}{7(2g+1)}$ \\ \vspace{4pt}
$F_{\nu}$ ($\nu_{\rm{m}}<\nu<\nu_{\rm{a}}<\nu_{\rm{c}}$) & $\frac{20-3k}{4(4-k)}$ & $\frac{113-22k}{24(4-k)}$ & $\frac{5(5g+8)}{14(2g+1)}$ \\ \vspace{4pt}
$F_{\nu}$ ($\nu_{\rm{m}}<\nu_{\rm{a}}<\nu<\nu_{\rm{c}}$) & $-\frac{3p(4-k)-12+5k}{4(4-k)}$ & $-\frac{p(73-14k)+3(7-2k)}{24(4-k)}$ & $-\frac{3p(5g+8)+7g}{14(2g+1)}$ \\ \vspace{2pt}
$F_{\nu}$ ($\nu_{\rm{m}}<\nu_{\rm{a}}<\nu_{\rm{c}}<\nu$) & $-\frac{3p-2}{4}$ & $-\frac{p(73-14k)+2(47-10k)}{24(4-k)}$ & $-\frac{3p(5g+8)+2(11g+12)}{14(2g+1)}$ \\
\hline
\end{tabular}
\end{center}
\end{table*}

In the remainder of this section we will discuss the observed light curves in terms of the reverse-forward shock and two-component jet model. 
In Table~\ref{tab:parscale} we give the temporal scalings of $F_{\nu,\rm{max}}$, $\nu_{\rm{m}}$, $\nu_{\rm{c}}$, and $F_{\nu}$ in various spectral regimes, for analytic forward and reverse shock models, to compare with the observed light curve slopes in Table~\ref{tab:lcdecay}. 
We note that the modeling results shown in Figures~\ref{fig:modelrevfor} and \ref{fig:modeldouble} are not formal fits, because of: (i) the extremely good quality of the data compared to the fairly simplified models applied here, which results in unreasonably high values for the fit statistic, and (ii) the number of parameters in, and complexity of, the two models. 
Therefore we can not statistically discriminate between the two models, but we discuss how well they describe the observed light curve features.

\subsubsection{Reverse-Forward Shock Model}\label{sec:modelrevfor}

\begin{figure*}
\begin{center}
\includegraphics[angle=-90,width=\textwidth]{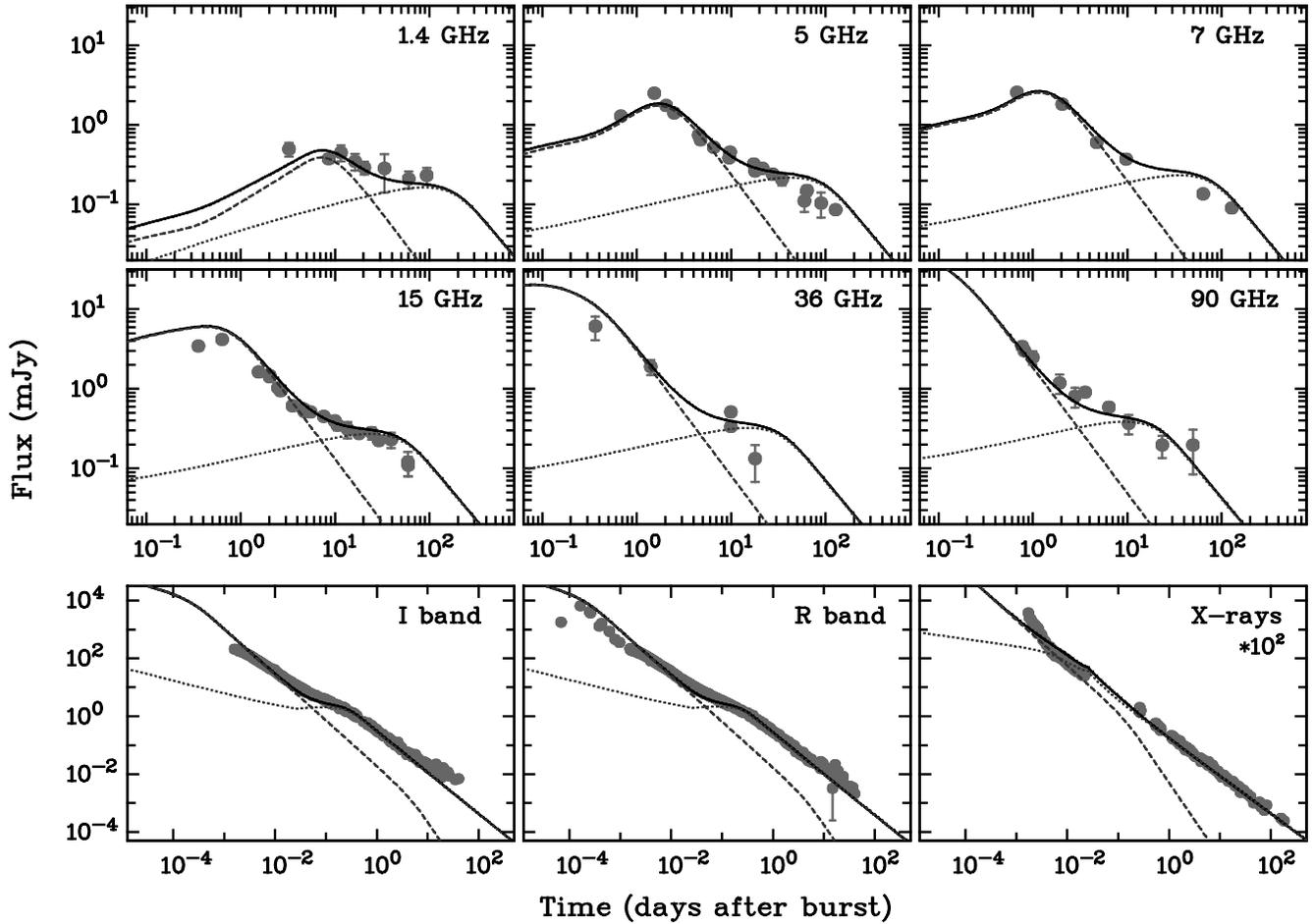}
\caption{Broadband modeling results for the reverse-forward shock model of all the available data at radio, optical, and X-ray frequencies \citep[Table~\ref{tab:radio} of this paper;][]{laskar2013,anderson2014,maselli2014,perley2014,vestrand2014}. 
The reverse shock is indicated with dashed lines, the forward shock with dotted lines, and the total flux with solid lines.}
\label{fig:modelrevfor}
\end{center}
\end{figure*}

In both the reverse-forward shock model and the two-component jet model the flattening of the optical light curves between 0.02 and 0.6~d is interpreted as $\nu_{\rm{m}}$ of a forward shock moving close to the observing bands; the transition to the final decay occurs when $\nu_{\rm{m}}$ has passed through a particular band. 
The flattening in the radio bands on the timescale of days to weeks, as well as the eventual light curve turnovers, are also interpreted by the passage of $\nu_{\rm{m}}$. 
Figures~\ref{fig:modelrevfor} and \ref{fig:modeldouble} show that when $\nu_{\rm{m}}$ is at optical frequencies, the peak flux $F_{\nu,\rm{max}}$ is a few mJy, while it is an order of magnitude lower when $\nu_{\rm{m}}$ passes through the radio bands. 
This is a clear indication that the ambient medium is not homogeneous, since $F_{\nu,\rm{max}}$ is then expected to be constant \citep{sari1998}. 
Therefore, we assume in our modeling that the ambient medium density is a power law with radius, $\rho=A\cdot R^{-k}$, where $k=0$ corresponds to a homogeneous medium and $k=2$ to a stellar wind with constant velocity. 
As can be seen in Table~\ref{tab:parscale}, $F_{\nu,\rm{max}}$ decreases in time for $k>0$. 
The cooling frequency decreases in time for a homogeneous medium but increases for a wind medium, while $\nu_{\rm{m}}$ is independent of the circumburst medium structure \citep[for the dependencies on all physical parameters we specifically use the equations in][]{vanderhorst2007}.

The evolution of $\nu_{\rm{m}}$, however, is not fast enough to account for the times at which it passes through the optical and radio bands, for which a temporal power-law index of $\sim-2$ is required. 
We have explored various possibilities to explain this behavior of $\nu_{\rm{m}}$, for instance the evolution after a jet break or a non-relativistic outflow. 
The light curve slopes, however, would then be significantly steeper than what has been observed, and these are, therefore, not viable explanations. 
We propose here that the fast evolution of $\nu_{\rm{m}}$ is caused by the temporal evolution of the microphysical parameters, as also suggested for other GRBs with well sampled light curves \citep[e.g.,][]{filgas2011}. 
While $F_{\nu,\rm{max}}$ and $\nu_{\rm{c}}$ do not depend on $\varepsilon_{\rm{e}}$, the peak frequency $\nu_{\rm{m}}\propto\varepsilon_{\rm{e}}^{2}$, and thus $\nu_{\rm{m}}\propto t^{-1.9}$ for a modest evolution of $\varepsilon_{\rm{e}}\propto t^{-0.2}$. 
We do not require any evolution of the other microphysical parameter $\varepsilon_{\rm{B}}$. 
Based on the late-time light curve slopes, the optical-to-X-ray spectra, and the temporal behavior of $F_{\nu,\rm{max}}$ and $\nu_{\rm{m}}$, we find that $k\simeq1.7$ and $p\simeq2.1$ describe the data well. 
This results in $F_{\nu,\rm{max}}\propto t^{-0.37}$ and $\nu_{\rm{c}}\propto t^{0.24}$. 
The light curves before the passage of $\nu_{\rm{m}}$ rise as $t^{0.26}$, 
and after the passage of $\nu_{\rm{m}}$ decay as $t^{-1.4}$, while above $\nu_{\rm{c}}$ they decay as $t^{-1.3}$.

For the reverse shock there are two possible evolution regimes, depending on the spread of outflow velocities in the shell behind the forward shock and the time it takes the reverse shock to cross this shell \citep{sari1995}. 
In the thin-shell or Newtonian case the outflow velocity spread is small, and the initially Newtonian reverse shock is still sub-relativistic once it has crossed the shell. 
If there is a large spread in the velocities, the shell spreads and the reverse shock becomes relativistic before crossing the entire shell, i.e., the thick-shell or relativistic case. 
From the temporal scalings in Table~\ref{tab:parscale} \citep[based on][]{kobayashi2000,chevalier2000,yi2013} we can derive that in the latter case the light curve slope for frequencies $\nu<\nu_{\rm{m,c}}$ is $-0.49$ for $k=1.7$, and $-2.1$ for $\nu_{\rm{m}}<\nu<\nu_{\rm{c}}$ and $p=2.1$. 
The slope for $\nu<\nu_{\rm{m,c}}$ is too shallow for the observed decay slopes ($\sim-1.2$ to $-1.4$), while for $\nu_{\rm{m}}<\nu<\nu_{\rm{c}}$ it is too steep. 
The latter is also true for $\nu_{\rm{m,c}}<\nu$ and $\nu_{\rm{c}}<\nu<\nu_{\rm{m}}$, and this large slope difference can not be accounted for by a moderate evolution of the microphysical parameters. 
Including self-absorption results in rising light curves for frequencies below $\nu_{\rm{a}}$, and can thus also not explain the observed light curves. 

For the thin-shell case the Lorentz factor of the ejecta is assumed to be a power law with radius, $\Gamma\propto R^{-g}$ \citep{meszaros1999}. 
{The result is that the temporal evolution of the characteristic frequencies and the peak flux, and therefore also the light curve slopes, depend on the power-law index $g$ \citep{kobayashi2000,zou2005}.} 
With $g$ as a free parameter we can describe the overall trends of the observed light curves fairly well, as shown in Figure~\ref{fig:modelrevfor}. 
We find that  $g\simeq5$, and that $\nu_{\rm{m}}>\nu_{\rm{a}}$ at early times and $\nu_{\rm{m}}=\nu_{\rm{a}}\simeq22$~GHz at $\sim0.4$~d. 
With this combination of parameters the radio light curves rise with a slope of $0.4$ for $\nu_{\rm{a}}<\nu_{\rm{m}}$ and $1.1$ for $\nu_{\rm{m}}<\nu_{\rm{a}}$, and the radio and optical light curves decay with a slope of $-1.6$. 
It is clear from Figure~\ref{fig:modelrevfor} that this gives a fairly good description of the radio light curves, even though it overestimates the peak at 15~GHz and underestimates the peak at 5 GHz, and it also follows the trend of the optical light curves after 0.004~d. 
However, the observed early-time optical light curves are over-estimated, because the observed flattening at $0.001-0.004$~d can not be reconstructed. 
Furthermore, the peak in the R-band light curve can not be explained in this model. 
This peak is so early that it could be caused by the reverse-forward shock system still building up, i.e., the peak of the light curve corresponds to the deceleration time scale. 
Alternatively, we note that for a significant evolution of $\varepsilon_{\rm{e}}$, i.e., $\varepsilon_{\rm{e}}\propto t^{-1}$, the R band model light curve does turn over at the peak, without very significantly affecting the later-time light curve or the results at other frequencies.

Despite the fact that the reverse-forward shock model describes the overall trends of the broadband light curves fairly well, we have also shown that there are some clear deviations when all of the available data are used in the modeling. 
Furthermore, the value of $g\simeq5$ is very high, and as also pointed out by other authors \citep{laskar2013,panaitescu2013} it is outside the range of theoretically allowed values, namely $3/2\le g\le7/2$ for a homogeneous medium \citep{kobayashi2000} and $1/2\le g\le3/2$ for a wind medium \citep{zou2005}. 
The lower bounds on $g$ are governed by the fact that the shell should lag behind the forward shock ($\Gamma\propto R^{-(3-k)/2}$), while the upper bound comes from the fact that the ejecta can not be quicker than in the relativistic case ($\Gamma\propto R^{-(7-2k)/2}$), so for $k\simeq1.7$ the allowed range is $0.65\le g\le1.8$. 
Values within this allowed range for $g$ result in significantly worse fits, i.e., much steeper light curve slopes (most notably an optical slope of $<-1.9$) and larger discrepancies at the peaks of the radio light curves. 
Because of these issues with the reverse-forward shock model we have explored a two-component jet model to fit the observed light curves.

\subsubsection{Two-Component Jet Model}\label{sec:modeldouble}

\begin{figure*}
\begin{center}
\includegraphics[angle=-90,width=\textwidth]{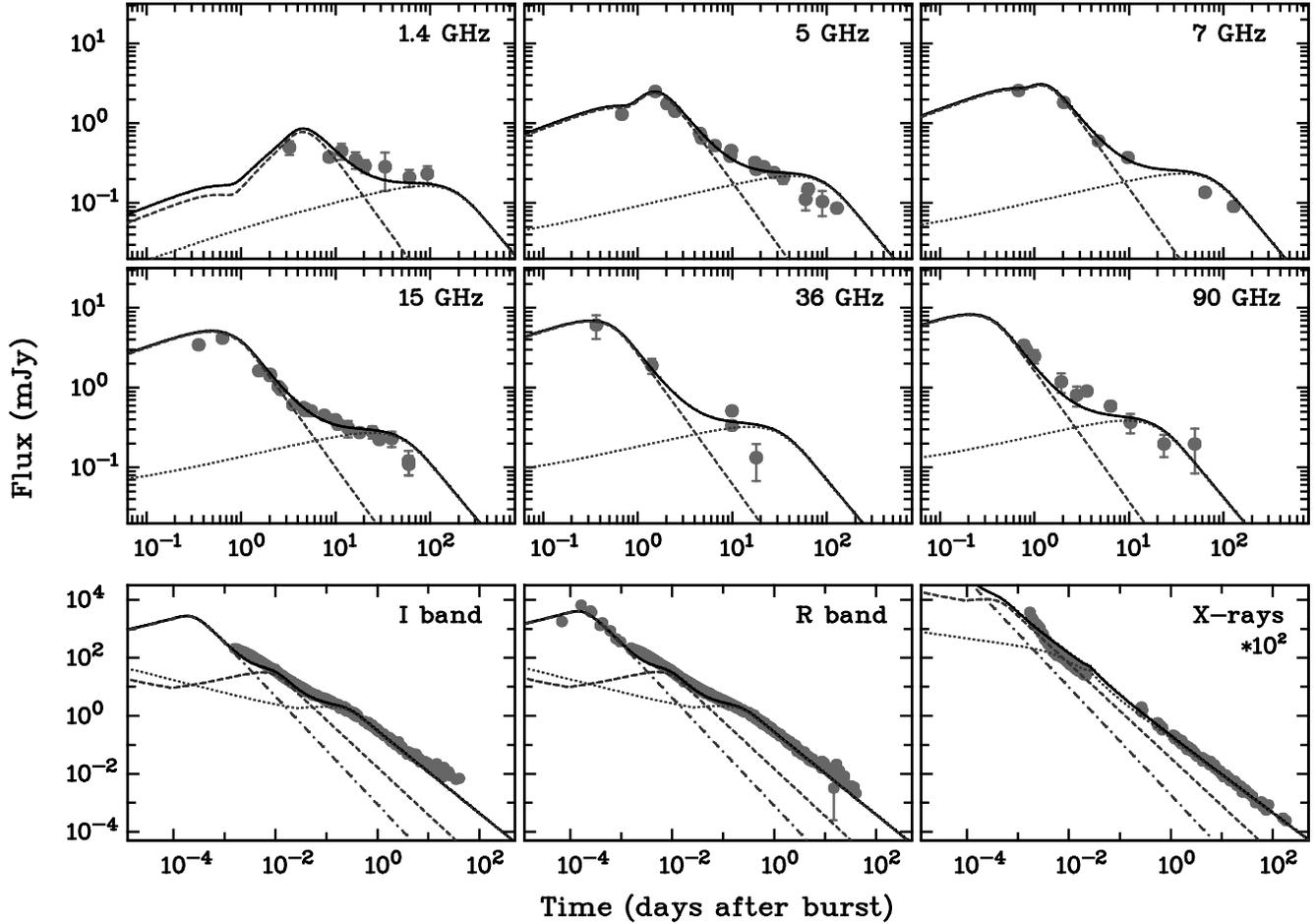}
\caption{Broadband modeling results for the two-component jet model at radio, optical, and X-ray frequencies. 
The narrow jet is indicated with dashed lines, the wide jet with dotted lines, the reverse shock with dash-dotted lines, and the total flux with solid lines.}
\label{fig:modeldouble}
\end{center}
\end{figure*}

The two-component jet model has been suggested to explain the broadband light curves and other observed phenomena in several GRBs \citep[e.g.][]{pedersen1998,frail2000a,berger2003,starling2005,racusin2008}. 
In this model there is a narrow uniform jet with a high Lorentz factor and a wider component with a lower Lorentz factor. 
Such a jet structure has been theoretically predicted in different models, e.g., a hydromagnetically driven neutron-rich jet \citep{vlahakis2003}, or a jet breakout from a progenitor star which results in a highly relativistic jet core surrounded by a moderately relativistic cocoon \citep{ramirezruiz2002}. 
Optical light curves for such jet structures have been calculated \citep{peng2005}, and using some combinations of physical parameters, the steep-flat-steep behavior observed in \grb can be retrieved. 
We applied a model consisting of two forward shocks to the broadband data of \grbnos, and as shown in Figure~\ref{fig:modeldouble}, this model can fit all the light curves well. 
The radio peak and early-time behavior, and the optical light curves between 0.004 and 0.02~d, are dominated by the narrow jet, while the late-time radio and optical light curves, and also the X-ray light curve, are dominated by the wide jet. 
The only feature that this model of two forward shocks can not explain is the very early-time behavior before 0.004~d in the R band, for which we invoke a reverse shock component.

In our two-component jet model the wide jet has the same parameters as the forward shock in the reverse-forward shock model of Section~\ref{sec:modelrevfor}. 
For this wide component we have constrained $F_{\nu,\rm{max,w}}$, $\nu_{\rm{m,w}}$, $\nu_{\rm{c,w}}$, $p=2.1$, $k=1.7$, and adopted $\varepsilon_{\rm{e,w}}\propto t^{-0.2}$, while $\nu_{\rm{a,w}}$ can not be determined. 
For the latter we can only put an upper limit of $\nu_{\rm{a,w}}<10^{9}$~Hz at 1~d for self-absorption to not affect the late-time radio light curve fits. 
Since the narrow and wide jet components are both moving through the same ambient medium, we assume that the density and its structure parameter $k$ are equal. 
We have not put any constraints on the other parameters for the narrow jet, since they can differ in energy; the microphysical parameters are also not necessarily the same for the two jet components. 
We find that $p=2.1$ also provides good fits for the narrow jet component, but we require a faster evolution of $\nu_{\rm{m,n}}\propto t^{-2.3}$, and therefore $\varepsilon_{\rm{e,n}}\propto t^{-0.4}$, $\nu_{\rm{a,n}}\propto t^{0.0}$ for $\nu_{\rm{a,n}}<\nu_{\rm{m,n}}$, and $\nu_{\rm{a,n}}\propto t^{-0.8}$ for $\nu_{\rm{m,n}}<\nu_{\rm{a,n}}$. 
We also find that $\nu_{\rm{m,n}}>\nu_{\rm{a,n}}$ at early times and $\nu_{\rm{m,n}}=\nu_{\rm{a,n}}\simeq9$~GHz at $\sim0.8$~d. 
The resulting light curve slopes are $0.5$ for $\nu<\nu_{\rm{a,n}}<\nu_{\rm{m,n}}$, $0.4$ for $\nu_{\rm{a,n}}<\nu<\nu_{\rm{m,n}}$, $0.7$ for $\nu_{\rm{m,n}}<\nu<\nu_{\rm{a,n}}$, and $-1.6$ for $\nu_{\rm{a,n}}<\nu_{\rm{m,n}}<\nu$. 
We can only put a lower limit on $\nu_{\rm{c,n}}$ for it to not affect the early-time optical light curves, namely $\nu_{\rm{c,n}}>10^{16}$~Hz at 1~d. 

\begin{figure}
\begin{center}
\includegraphics[angle=-90,width=\columnwidth]{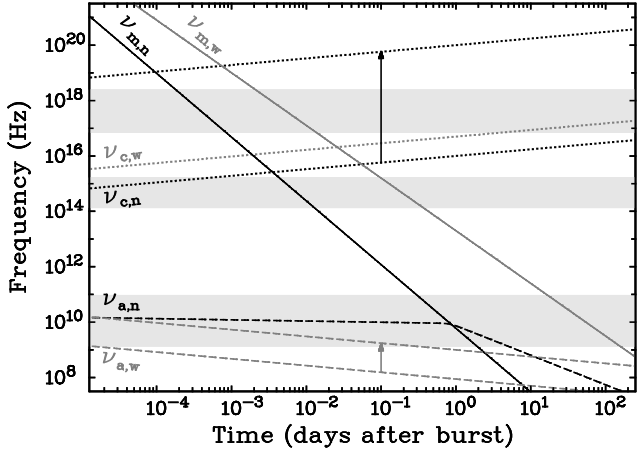}
\caption{Evolution of the characteristic frequencies in the two-component jet model for \grbnos. 
The black lines are for the narrow jet component, and the grey lines for the wide component; 
the solid lines are for $\nu_{\rm{m}}$, the dashed lines for $\nu_{\rm{a}}$, 
and the dotted lines for $\nu_{\rm{c}}$. 
The lower and upper limits for $\nu_{\rm{a,w}}$ and $\nu_{\rm{c,n}}$ are connected by arrows. 
The light grey bands indicate the X-ray, near-infrared/optical/UV, and radio observing bands which the characteristic frequencies move through.}
\label{fig:parsdouble}
\end{center}
\end{figure}

The evolution of the characteristic frequencies of both jet components is shown in Figure~\ref{fig:parsdouble}, illustrating when several of these parameters move through the observing bands. 
We can not determine $\nu_{\rm{a,w}}$ of the wide jet component nor $\nu_{\rm{c,n}}$ of the narrow jet component, but we included the constraint that the ambient medium density is the same for both components, which means that there is still one free parameter. 
Given the constraint on the density, and the aforementioned limits on $\nu_{\rm{a,w}}$ and $\nu_{\rm{c,n}}$, we can determine allowed parameter ranges, which we give in Table~\ref{tab:params}. 
The table shows that the allowed parameter ranges include values for $\varepsilon_{\rm{e}}$ and $\varepsilon_{\rm{B}}$ that are larger than 1 for both jet components. 
These two parameters are fractions which are supposed to be smaller than 1, and in fact $\varepsilon_{\rm{e}}+\varepsilon_{\rm{B}}<1$ would be expected. 
If we take the values for $\varepsilon_{\rm{e}}$ at 0.001~d, the earliest time at which the narrow jet component is significantly contributing to the total flux, the lowest values for this sum are $\varepsilon_{\rm{e,n}}+\varepsilon_{\rm{B,n}}=3.0$ for $\nu_{\rm{c,n}}=1\times10^{17}$~Hz, and $\varepsilon_{\rm{e,w}}+\varepsilon_{\rm{B,w}}=6.6$ for $\nu_{\rm{c,n}}=3\times10^{16}$~Hz. 
These parameter values, however, are determined assuming that all the electrons that are swept up by the shocks are accelerated into the power-law energy distribution that produces the synchrotron radiation, while this is in fact only true for a fraction $\xi$ of the electrons. 
\citet{eichler2005} have shown that the observed emission does not change for the following scalings: $\varepsilon_{\rm{e}}\rightarrow\xi\varepsilon_{\rm{e}}$, $\varepsilon_{\rm{B}}\rightarrow\xi\varepsilon_{\rm{B}}$, $E\rightarrow E/\xi$, $\rho\rightarrow\rho/\xi$. 
To fulfill the requirement that $\varepsilon_{\rm{e}}+\varepsilon_{\rm{B}}<1$ for both jet components, $\xi<0.15$ is necessary (assuming that $\xi$ is independent of time or the shock Lorentz factor). 
This value is an important input for theoretical studies and simulations of particle acceleration in relativistic shocks.

In Table~\ref{tab:params} we give the values and time evolution for the radii $R$ and Lorentz factors $\Gamma$ of the two shocks.
The Lorentz factor of the narrow jet component is larger than the one of the wide component, which is indeed expected from theoretical studies and simulations. 
Both shocks are still extremely relativistic at 1~d, and their radii are large, which is mainly due to the low density. 
From the radii and Lorentz factors in Table~\ref{tab:params} we can estimate upper limits on the image radius at the moment of our EVN observation by assuming a spherical model \citep{granot2002}. 
At 6.55~d the narrow jet component has a size of $(2-8)\times10^{17}$~cm, and the wide jet component $(0.6-1)\times10^{18}$~cm, which are both smaller than the EVN upper limit on the radius of $5\times10^{18}$~cm. 

The value for $A$ in Table~\ref{tab:params} corresponds to a density of $7\times10^{-6}-9\times10^{-4}$~g/cm$^{3}$ at 1~pc. 
Since $k=1.7$ is close to the density structure of a stellar wind with a constant velocity ($k=2$), we estimate the mass-loss rate that would result in the derived density range at 1~pc: $2\times10^{-9}-3\times10^{-7}$~M$_{\odot}$/yr, assuming a typical wind velocity of $10^3$~km/s.
This kind of mass-loss rate is very low for typical Wolf-Rayet stars, usually assumed to be the progenitors of GRBs and with typical mass-loss rates of $10^{-5}$~M$_{\odot}$/yr. 
However, if the metallicity is significantly lower than solar metallicity, i.e., $<10^{-3}$, the mass-loss rates for Wolf-Rayet stars can be as low as $10^{-7}-10^{-8}$~M$_{\odot}$/yr \citep[depending on the type of Wolf-Rayet star;][]{vink2005}. 
The inferred mass-loss rates are also characteristic for late-type O stars (O6.5 to O9.5) in a wide range of metallicities \citep{vink2001}. 
We note that it has been suggested that O-emission stars that are rapidly rotating and have low metallicity, are indeed possible progenitors for GRBs \citep{woosley2006}. 

Another effect of the low density is that we have not observed a jet break in the light curves of \grbnos. 
The jet-break time $t_{\rm{j}}$ can be estimated by assuming that the jet opening angle $\theta$ is equal to $\Gamma^{-1}$, which implies that $t_{\rm{j,n}}=(6\times10^2-4\times10^3)\cdot\theta_{\rm{-1,n}}^{3.5}$~d and $t_{\rm{j,w}}=(6-1\times10^3)\cdot\theta_{\rm{-1,w}}^{3.5}$~d, with $\theta_{\rm{-1}}=\theta/0.1$~rad. 
From the lack of any jet break in the light curves we deduce that $t_{\rm{j,n}}>20$~d, since the narrow jet does not contribute to the total flux anymore after this time, and $t_{\rm{j,w}}>120$~d, the latest reported detection of the source; and thus $\theta_{\rm{n}}>1\degr$ and $\theta_{\rm{w}}>3\degr$. 
Based on these lower limits on the opening angles and the isotropic equivalent energies given in Table~\ref{tab:params}, we derive the ranges for the collimation corrected energies of $7\times10^{49}<E_{\rm{j,n}}<3\times10^{54}$~erg and $8\times10^{49}<E_{\rm{j,w}}<6\times10^{52}$~erg.

For the reverse shock that gives rise to the early optical light curve peak in the two-component jet model we can not constrain the physical parameters well. 
In Figure~\ref{fig:modeldouble} we show the model light curve for a thin-shell reverse shock with $g=1.8$, in which the light curve peak is caused by the passage of $\nu_{\rm{a}}$ for $\nu_{\rm{m}}<\nu_{\rm{a}}$. 
The correct light curve slopes can be obtained by this ordering of the characteristic frequencies, but the rising part and the peak of the optical light curve can also be caused by the end of the passage of the reverse shock through the shell. 
Due to the lack of observations at other frequencies at similarly early times the parameters of the reverse shock can not be determined.

We conclude that the two-component jet model is a good alternative for the reverse-forward shock model proposed by other authors, in terms of describing the broadband light curves. 
We would like to point out, however, that we assumed that $\varepsilon_{\rm{e}}$ and $\varepsilon_{\rm{B}}$ are not the same for the wide and narrow jet, and we find the ratios $\varepsilon_{\rm{B,n}}/\varepsilon_{\rm{B,w}}=0.02-4$ and $\varepsilon_{\rm{e,n}}/\varepsilon_{\rm{e,w}}=(0.08-0.11)\cdot t_{\rm{d}}^{-0.2}$. 
The range for $\varepsilon_{\rm{e,n}}/\varepsilon_{\rm{e,w}}$ is significantly smaller than the range for $\varepsilon_{\rm{B,n}}/\varepsilon_{\rm{B,w}}$, but $\varepsilon_{\rm{B,n}}=\varepsilon_{\rm{B,w}}$ is true for $\nu_{\rm{c,n}}\simeq1\times10^{17}$~Hz. 
The ratio for $\varepsilon_{\rm{e}}$ is time-dependent, and $\varepsilon_{\rm{e,n}}=\varepsilon_{\rm{e,w}}$ is fulfilled at $\sim10^{-5}$~d, which is in the first second after the GRB onset. 
Regarding $\varepsilon_{\rm{e}}$ one expects that this parameter is the same for two shocks with the same Lorentz factor moving into the same medium, and that this is also true for its temporal evolution. 
When calculating $\varepsilon_{\rm{e,n}}/\varepsilon_{\rm{e,w}}$ for the same Lorentz factor, this ratio is still significantly deviating from $1$, in contrast with what is expected based on theoretical grounds, while light curves for $\varepsilon_{\rm{e,n}}=\varepsilon_{\rm{e,w}}$ result in significantly worse fits. 
We conclude that both the reverse-forward shock model and the two-component jet model have an issue in the sense that one of the parameters that provide the best description of the broadband light curves is outside the range of theoretically allowed or expected values.

\begin{table}
\begin{center}
\caption{Physical parameters for the two-component jet model, with the fraction of electrons participating in a relativistic power-law energy distribution set to $\xi=1$, the density $\rho=A\cdot R^{-1.7}$, and $t_{\rm{d}}$ the time in days.}
\label{tab:params}
\renewcommand{\arraystretch}{1.1}
\begin{tabular}{|l|c|c|}
\hline
Parameter & Narrow Jet & Wide Jet  \\
\hline
$E_{\rm{iso}}$ (erg) & $3\times10^{53}-3\times10^{54}$ & $8\times10^{51}-6\times10^{52}$ \\
$A$ (g/cm$^{1.3}$) & $3\times10^{2}-4\times10^{4}$ & $3\times10^{2}-4\times10^{4}$ \\
$\varepsilon_{\rm{B}}$ & $1\times10^{-4}-1\times10^{1}$ & $8\times10^{-3}-3$ \\
$\varepsilon_{\rm{e}}$ & $(0.08-0.8)\cdot t_{\rm{d}}^{-0.4}$ & $(1-7)\cdot t_{\rm{d}}^{-0.2}$ \\
\hline
$R$ (cm) & $(0.9-3)\times10^{19}\cdot t_{\rm{d}}^{0.43}$ & $(0.07-2)\times10^{19}\cdot t_{\rm{d}}^{0.43}$  \\
$\Gamma$ & $(0.6-1)\times10^{2}\cdot t_{\rm{d}}^{-0.28}$ & $(2-8)\times10^{1}\cdot t_{\rm{d}}^{-0.28}$ \\
\hline
\end{tabular}
\end{center} 
\end{table}

\section{Short Timescale Variability}\label{sec:shvar}

\begin{figure}
\begin{center}
\includegraphics[angle=-90,width=\columnwidth]{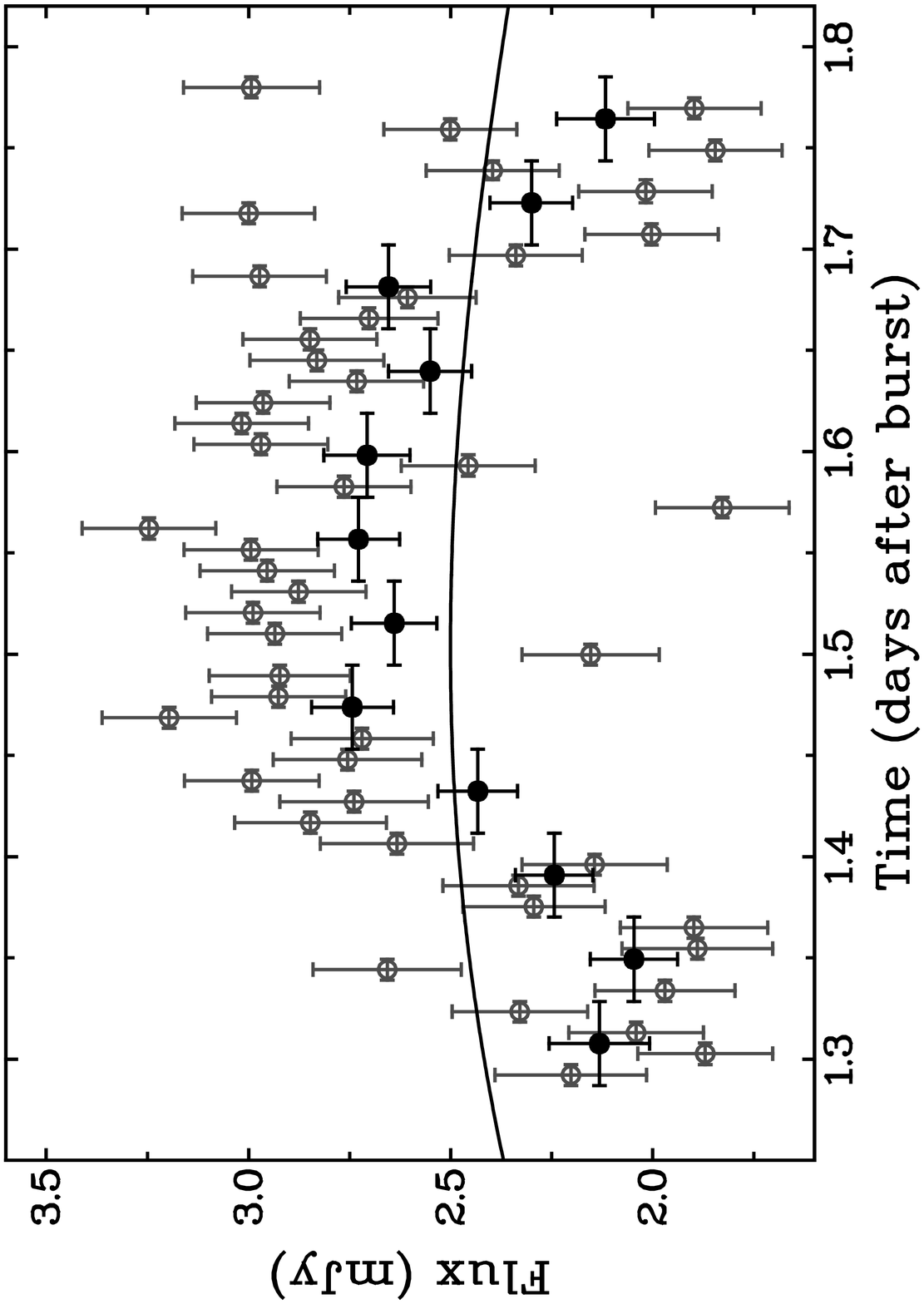} \\ \vspace{0.2cm}
\includegraphics[angle=-90,width=\columnwidth]{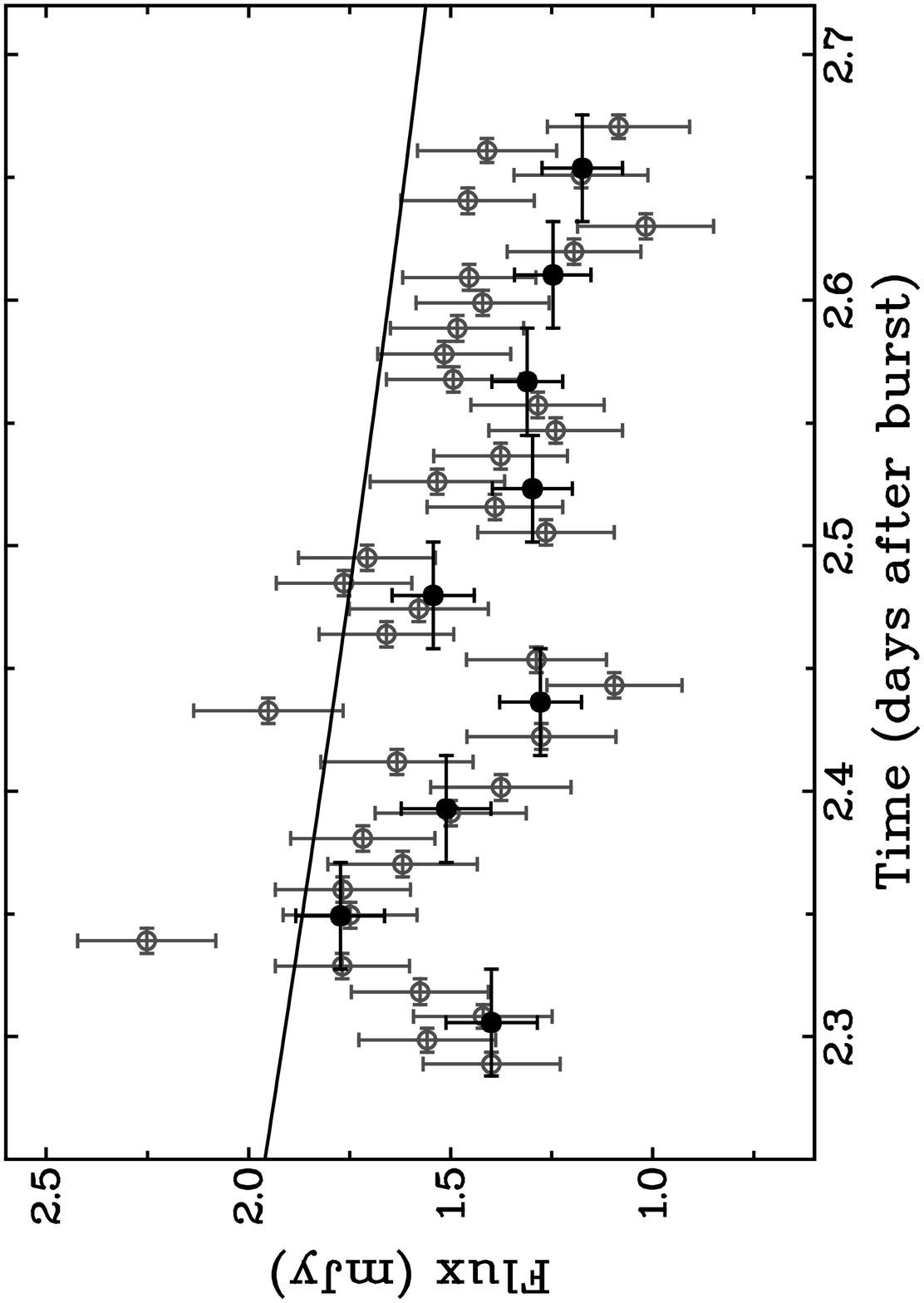}
\caption{Detailed light curve at 4.8~GHz of the 12-hour observation on April 28.6$-$29.1 (top panel) and the 9.4-hour observation on April 29.6$-$30.0, at a time resolution of one hour (solid black symbols) and 15 minutes (open grey symbols). 
The solid line shows the two-component jet model presented in Figure~\ref{fig:modeldouble}.}
\label{fig:radiolcsfine}
\end{center}
\end{figure}

The first two WSRT observations of \grb at 4.8~GHz were 12 and 9.4 hours in duration, respectively. 
Since the source was so radio bright in the first few days, and we had continuous observations at one frequency for so many hours (while we were doing frequency switching between 4.8 and 1.4~GHz in following epochs), we had a sufficiently high signal-to-noise ratio to determine the flux evolution within these two observations. 
Figure~\ref{fig:radiolcsfine} shows the light curves for the first two epochs with a time resolution of 15~minutes (grey open symbols) and 1~hour (black solid symbols; see Section~\ref{sec:wsrt} for the analysis details). 
From this figure it is clear that there are significant fluctuations in the observed flux. 
The first observation is during the peak of the light curve, and it also shows the rise, peak and decay. 
However, the rise and decay we observe seem to be significantly steeper than what would be expected from modeling, while the peak is broader than expected. 
The second observation is during the decay of the light curve, but shows fluctuations around an average decaying behavior. 
These kind of flux variations are not expected to be intrinsic to the source; they are most likely caused by interstellar scintillation \citep[ISS;][]{rickett1990,goodman1997}. 
The effects of ISS have been observed in several GRBs over timescales of days to weeks \citep[e.g.][]{frail1997,frail2000b}. 
In one GRB intraday variability during long observations, similar to what we observe in \grbnos, has also been found \citep[GRB\,070125;][]{chandra2008}. 
We will discuss if ISS can indeed explain the observed radio variability in \grbnos.

ISS is caused by propagation effects in the interstellar medium due to fluctuations in the density of free electrons. 
The scintillation strength and timescale depend on the observing frequency and the angular size of the source compared to characteristic scintillation angular scales. 
The observing frequency determines if the scattering is in the weak or strong regime, where in the strong scattering regime both refractive and diffractive scintillation can play a role. 
To estimate the transition frequency between weak and strong ISS, and the scattering measure, we adopt the NE2001 model for the distribution of free electrons in our galaxy \citep{cordes2002}. 
We note that this model is rather uncertain for sightlines off the galactic plane, and should thus be interpreted with caution, but it still provides decent estimates for the scintillation parameters. 
For \grb the galactic longitude and latitude are $l=206.5^{\circ}$ and $b=72.5^{\circ}$, respectively, which result in a transition frequency $\nu_0=6.77$~GHz between weak and strong scattering, and a scattering measure $SM=1.04\times10^{-4}$~kpc/m$^{20/3}$. 
This value for $\nu_0$ implies that our WSRT measurements are possibly affected by strong scattering, while the observations at higher frequencies are in the weak scattering regime. 
The ISS angular scales are proportional to the angular size of the first Fresnel zone $\theta_{\rm{F}0}=2.1\times10^4\,SM^{0.6}\,\nu_0^{-2.2}\,=\,1.3\,\mu\rm{as}$ \citep{walker1998}. 
At the redshift $z=0.34$ of \grb an angular size of 1~$\mu$as corresponds to a physical size of $1.5\times10^{16}$~cm, which means that $\theta_{\rm{F}0}$ corresponds to a source size of $1.9\times10^{16}$~cm.

Since our intraday variability measurements are in the strong scattering regime, we have determined the angular scales $\theta$, variability time scales $t$, and modulation indices $m$ for refractive and diffractive scintillation. 
For an observing frequency $\nu=4.8$~GHz the refractive scintillation parameters are $\theta_{\rm{r}}=\theta_{\rm{F0}}(\nu/\nu_0)^{-11/5}=2.7\,\mu\rm{as}$, $t_{\rm{r}}=2(\nu/\nu_0)^{-11/5}=4.3\,\rm{hours}$, and $m_{\rm{r}}=(\nu/\nu_0)^{17/30}=0.82$. For diffractive scintillation $\theta_{\rm{d}}=\theta_{\rm{F0}}(\nu/\nu_0)^{6/5}=0.84\,\mu\rm{as}$, $t_{\rm{d}}=2(\nu/\nu_0)^{6/5}=1.3\,\rm{hours}$, and $m_{\rm{d}}=1$. 
Diffractive scintillation is a narrowband phenomenon, and for it to have a maximum effect the observing bandwidth should be less than $\Delta\nu_{\rm{d}}=\nu(\nu/\nu_0)^{17/5}=1.4\,\rm{GHz}$, which is indeed the case for our WSRT observations with a bandwidth of 160~MHz. 
Based on the light curves in Figure~\ref{fig:radiolcsfine} the flux variations are occurring at timescales of an hour to a few hours, which implies that both diffractive and refractive scintillation could be playing a role. 
This puts constraints on the size of the emission in the first couple of days after the GRB onset, since $\theta_{\rm{d}}$ and $\theta_{\rm{r}}$ correspond to physical source sizes of $1.3\times10^{16}\,\rm{cm}$ and $4.0\times10^{16}\,\rm{cm}$ for diffractive and refractive scintillation, respectively. 
Once the source size $\theta_{\rm{s}}$ becomes larger than $\theta_{\rm{d}}$ or $\theta_{\rm{r}}$, the variability timescales will increase with a factor $\theta_{\rm{s}}/\theta_{\rm{d}}$ or $\theta_{\rm{s}}/\theta_{\rm{r}}$, respectively, while the modulation indices will decrease with a factor $(\theta_{\rm{s}}/\theta_{\rm{d}})^{-1}$ for diffractive scintillation and $(\theta_{\rm{s}}/\theta_{\rm{r}})^{-7/6}$ for refractive scintillation. 

Figure~\ref{fig:radiolcsfine} shows that the flux modulations are largest in the first WSRT epoch. 
We estimate the maximum observed modulation index by determining the largest deviation from the model fit in the 1-hour data, which implies a modulation index $m=0.12$. 
Based on the jet radii and Lorentz factors inferred from our modeling in Section~\ref{sec:modeldouble}, we estimate upper limits on the image radii of both jet components by assuming a spherical model \citep{granot2002}: $(2-4)\times10^{17}$~cm for the narrow jet and $(0.6-3)\times10^{17}$~cm for the wide jet component. 
These inferred image radii are larger than the source sizes corresponding to $\theta_{\rm{d}}$ and $\theta_{\rm{r}}$, 
which implies that the minimum modulation indices are $m_{\rm{d}}=0.04-0.06$ and $m_{\rm{r}}=0.06-0.12$ for the narrow jet, and $m_{\rm{d}}=0.05-0.21$ and $m_{\rm{r}}=0.09-0.54$ for the wide jet component. 
These modulations indices are consistent with the observed modulation index $m=0.12$. 
The inferred scintillation timescales are $22-39$~h for the narrow jet, and $6-28$~h for the wide jet component. 
While these timescales are long compared to the scintillation behavior we observe, in particular the ones for the narrow jet, the modulation indices we inferred are also low compared to the observed value, and both of these discrepancies can be resolved if one takes into account that we are dealing with jets instead of a spherical outflow. 
However, we have already noted that the estimates for $\nu_0$ and $SM$ are quite uncertain far away from the galactic plane, and it has also been shown for quasars displaying intraday variability that the scattering medium can be significantly closer than what is usually assumed \citep{dennettthorpe2002,bignall2006,macquart2007}. 
Given our estimates we conclude that our observed flux modulations are consistent with both diffractive and refractive ISS, but due to the uncertainties in the properties of the scattering medium we can not put any further constraints on the size or opening angle of the jet.

For completeness, we have also calculated the possible effect of weak scintillation on observations at higher frequencies, in particular for the well-sampled light curve at 15~GHz. 
The angular scale is in this case $\theta_{\rm{w}}=\theta_{\rm{F0}}(\nu/\nu_0)^{-1/2}=0.86\,\mu\rm{as}$ and the variability timescale is $t_{\rm{w}}=2(\nu/\nu_0)^{-1/2}=1.3\,\rm{hours}$, both comparable to the values for diffractive scintillation at 4.8~GHz. 
The modulation index, however, is significantly smaller: $m_{\rm{w}}=0.33$, and decreases by a factor $(\theta_{\rm{s}}/\theta_{\rm{w}})^{-7/6}$ once $\theta_{\rm{s}}>\theta_{\rm{w}}$. 
The fact that no significant flux variations are observed in the 15~GHz light curve is consistent with the short variability timescale and the low modulation index.

\section{Polarization}\label{sec:pola}

Measuring polarization in GRBs, or any other astrophysical source, is important for putting constraints on the magnetic field structure in the emission regions. 
Variable optical linear polarization at a few percent level has been found at a timescale of hours to days after the GRB onset \citep[e.g.,][]{covino1999,wijers1999}, but due to challenges of observing these low levels there are only a few well sampled polarization curves \citep{greiner2003,wiersema2012}. 
Recently, optical observations in the first minutes of two GRBs have revealed linear polarizations of $10\%$ \citep{steele2009} to $28\%$ \citep{mundell2013}. 
At those early times the reverse shock can contribute significantly to the observed emission. 
Since the reverse shock probes the GRB outflow, this suggests that the magnetic field in the jet is uniform over large scales. 

Searches for polarization at radio frequencies have been undertaken, but have so far been unsuccessful. 
The most stringent constraints have been obtained for GRB\,030329, with a linear polarization limit $<1.0\%$ at 7.7~days \citep{taylor2004}, and limits of $1.8\%$ and $4.7\%$ at 3 and 7 months, respectively \citep{taylor2005}. 
All these observations were performed at late times when the forward shock was producing the observed emission. 
The polarization during a radio flare, and thus possible reverse shock emission, has been constrained for three GRBs \citep{granot2005}. 
For GRB~990123 and GRB~020405 the limits on the linear and circular polarization were larger than $10\%$ at $\sim1.2$~days. 
The best limits were obtained by combining two observations of GRB\,991216, at 1.5 and 2.7 days, to obtain a linear polarization $P_{\rm{L}}<7\%$ and a circular polarization $P_{\rm{C}}<9\%$. 
Our polarization limits for \grb are obtained at similar times: $P_{\rm{L}}<3.9\%$ and  $P_{\rm{C}}<2.7\%$ at 1.5 days, and $P_{\rm{L}}<7.5\%$ and  $P_{\rm{C}}<5.7\%$ at 2.5 days. 

The interpretation of our polarization limits depends on the nature of the radio peak, i.e., whether it is reverse shock emission or produced by the narrow jet in a two-component jet model. 
A further complication is that there are no optical polarization measurements for \grb at the time of the optical peak \citep[and an upper limit $<3\%$ of the optical linear polarization from 0.16 to 0.42~d, when forward shock emission is dominating in both models;][]{itoh2013}. 
Our polarization limits of a few percent are lower than the optical polarization levels observed at very early times for two other GRBs \citep{steele2009,mundell2013}. 
If the optical flash in \grb were polarized at the tens of percent level, this would have provided important information on the size scale over which the magnetic field in the jet is uniform in the reverse shock scenario \citep{granot2003}. 
Because of relativistic beaming we only see emission from a region with an angle $\sim1/\Gamma$ around our line of sight. 
At the time of the optical flash $\Gamma$ is typically of the order of several hundreds, while at the time of the radio flare it has usually decelerated to a few tens. 
This implies that the emission region we are observing has increased from $<0.01$ to $\sim0.1$~rad, and while the magnetic field can be uniform over the former angular scale, this is not necessarily the case for the latter angular scale. 
This can lead to a significant decrease in radio polarization from the reverse shock compared to the optical polarization. 

These considerations are true for a reverse shock interpretation of the radio flare in \grbnos, but we have shown in Section~\ref{sec:model} that a two-component jet model provides a good alternative to describe the data. 
Our linear polarization limits, in particular the ones at the first WSRT epoch, are close to the linear polarization levels measured for optical forward shock emission in other GRBs, although our circular polarization limits are significantly higher than the optical levels \citep{wiersema2014}. 
An important effect to take into account when comparing radio to optical polarization is synchrotron self-absorption, which we have shown plays a role at radio frequencies in \grb (Section~\ref{sec:model}). 
This can suppress the linear polarization \citep{toma2008}, but can in fact enhance the circular polarization to higher levels than at optical frequencies \citep{matsumiya2003}. 
The latter can reach levels of $\sim1\%$, which is still below but close to our observed circular polarization constraints in the first epoch. 
We note that propagation through the media between the source and us can cause depolarization, but it has been argued that this effect is not very large for GRBs at radio frequencies \citep{granot2005}.
To conclude, while the WSRT polarization limits for \grb are among the lowest radio polarization limits to date, due to the lack of optical polarization detections we can not put robust constraints on jet or emission models, especially when one takes relativistic and self-absorption effects into account. 
Even deeper radio polarization measurements, and especially combined with optical polarization observations, will be necessary to constrain jet models in other GRBs.

\section{Conclusions}\label{sec:concl}

\grb was a record-breaking GRB in many respects, and its broadband follow-up from GHz radio frequencies to GeV gamma-ray energies has resulted in very well sampled light curves. 
In this paper we have presented radio observations with the WSRT at 1.4 and 4.8~GHz, significantly enhancing the temporal coverage at these two frequencies. 
We have combined our WSRT observations with data published in the literature and performed broadband modeling. 
We have shown that the reverse-forward shock model put forward by other authors can not fit all the light curves well, plus the obtained dependence of the outflow Lorentz factor on radius is not physical. 
As an alternative we have shown that the addition of a second jet component provides a good description of the light curves from radio to X-ray frequencies, in particular that the very early steep decay and subsequent flattening in the optical light curve can be described well by adding the extra free parameters of a second forward shock emission component. 
In this model only the very early optical peak originates in the reverse shock, while the rest of the optical emission, and also the radio and X-ray emission, are produced by a narrow fast jet surrounded by a slower and wider jet component. 
We can not determine which one of the two models is statistically better, but we can draw conclusions on the physics of the jet and its surroundings that are true for both models. 
We have put constraints on the physical parameters, and found that the density is very low and structured like a stellar wind. 
The low density indicates a very low mass-loss rate from the progenitor star, which implies either a low-metallicity ($<10^{-3}$ of solar metallicity), nitrogen-rich Wolf-Rayet star; or a rapidly rotating, low-metallicity O star. 
We have also determined the microphysical parameters describing the energetics of the electrons and magnetic field. 
To explain the fast evolution of the spectral peak frequency, we have invoked a moderate temporal evolution of $\varepsilon_{\rm{e}}$. 
Furthermore, we find that the fraction of electrons participating in a relativistic power-law energy distribution is $<15\%$. 
We note that one issue with the two-component jet model is that the temporal evolution of $\varepsilon_{\rm{e}}$ is slightly different for the narrow and wide jet components, and that they are only equal to each other at $\sim1$~s after the GRB onset.

Besides radio flux density measurements we have also performed VLBI observations to constrain the source size at 6.55~d. 
Unfortunately the source became too faint for VLBI observations at later times, when measuring the source size with this technique would have been feasible, but we did obtain the most accurate localization of this GRB. 
Because of the long observations at 4.8~GHz and the brightness of the source we were able to study intraday variability within the first days after the GRB onset. 
In particular the observation at $\sim1.5$~d showed fast variations which were not intrinsic to the source, and most likely caused by strong ISS. 
We showed that this is indeed a plausible explanation by comparing the source image size inferred from broadband modeling with the characteristic angular scales for ISS. 

Finally, we have presented some of the most constraining upper limits of radio polarization. 
These limits, of only a few percent on both linear and circular polarization, are at the peak of the 4.8~GHz radio emission. 
If one interprets this peak as emission from the reverse shock, these would be the deepest reverse shock radio polarization measurements. 
In our modeling work, however, we have shown that the radio peak can also be caused by the narrow core component of the jet, and although these polarization limits are still among the lowest ones to date (except for GRB\,030329), a non-detection of radio polarization at a few percent level is not unexpected (even for reverse shock emission). 
Pushing these limits further down in future GRB observations will allow us to put constraints on jet models, in particular the role and structure of magnetic fields in the jet and in the shocks producing the emission.

\section*{Acknowledgments}
AJvdH would like to thank Alex de Koter, Stan Woosley and Enrico Ramirez-Ruiz for helpful discussions. 
We greatly appreciate the support from the WSRT staff in their help with scheduling and obtaining the observations presented in this paper. 
The WSRT is operated by ASTRON (Netherlands Institute for Radio Astronomy) with support from the Netherlands foundation for Scientific Research. 
The EVN (\texttt{http://www.evlbi.org}) is a joint facility of European, Chinese, South African, and other radio astronomy institutes funded by their national research councils. 
The research leading to these results has received funding from the European Commission Seventh Framework Programme (FP/2007-2013) under grant agreement No. 283393 (RadioNet3). 
AIPS is produced and maintained by the National Radio Astronomy Observatory, a facility of the National Science Foundation operated under cooperative agreement by Associated Universities, Inc. 
AJvdH, RAMJW and AR acknowledge the support of the European Research Council Advanced Investigator Grant no. 247295. 
KW acknowledges support from STFC. 
RLCS is supported by a Royal Society Fellowship. 
PAC is supported by Australian Research Council grant DP120102393. 
GEA and RPF acknowledge the support of the European Research Council Advanced Investigator Grant no. 267697. 

\bibliographystyle{mn2e}
\bibliography{references}

\begin{thebibliography}{78}
\expandafter\ifx\csname natexlab\endcsname\relax\def\natexlab#1{#1}\fi

\bibitem[{{Ackermann} {et~al}\mbox{.}(2014){Ackermann}, {Ajello}, {Asano},
  {Atwood}, {Axelsson}, {Baldini}, {Ballet}, {Barbiellini}, {Baring},
  {Bastieri}, {Bechtol}, {Bellazzini}, {Bissaldi}, {Bonamente}, {Bregeon},
  {Brigida}, {Bruel}, {Buehler}, {Burgess}, {Buson}, {Caliandro}, {Cameron},
  {Caraveo}, {Cecchi}, {Chaplin}, {Charles}, {Chekhtman}, {Cheung}, {Chiang},
  {Chiaro}, {Ciprini}, {Claus}, {Cleveland}, {Cohen-Tanugi}, {Collazzi},
  {Cominsky}, {Connaughton}, {Conrad}, {Cutini}, {D'Ammando}, {de Angelis},
  {DeKlotz}, {de Palma}, {Dermer}, {Desiante}, {Diekmann}, {Di Venere},
  {Drell}, {Drlica-Wagner}, {Favuzzi}, {Fegan}, {Ferrara}, {Finke},
  {Fitzpatrick}, {Focke}, {Franckowiak}, {Fukazawa}, {Funk}, {Fusco},
  {Gargano}, {Gehrels}, {Germani}, {Gibby}, {Giglietto}, {Giles}, {Giordano},
  {Giroletti}, {Godfrey}, {Granot}, {Grenier}, {Grove}, {Gruber}, {Guiriec},
  {Hadasch}, {Hanabata}, {Harding}, {Hayashida}, {Hays}, {Horan}, {Hughes},
  {Inoue}, {Jogler}, {J{\'o}hannesson}, {Johnson}, {Kawano}, {Kn{\"o}dlseder},
  {Kocevski}, {Kuss}, {Lande}, {Larsson}, {Latronico}, {Longo}, {Loparco},
  {Lovellette}, {Lubrano}, {Mayer}, {Mazziotta}, {McEnery}, {Michelson},
  {Mizuno}, {Moiseev}, {Monzani}, {Moretti}, {Morselli}, {Moskalenko},
  {Murgia}, {Nemmen}, {Nuss}, {Ohno}, {Ohsugi}, {Okumura}, {Omodei}, {Orienti},
  {Paneque}, {Pelassa}, {Perkins}, {Pesce-Rollins}, {Petrosian}, {Piron},
  {Pivato}, {Porter}, {Racusin}, {Rain{\`o}}, {Rando}, {Razzano}, {Razzaque},
  {Reimer}, {Reimer}, {Ritz}, {Roth}, {Ryde}, {Sartori}, {Parkinson},
  {Scargle}, {Schulz}, {Sgr{\`o}}, {Siskind}, {Sonbas}, {Spandre}, {Spinelli},
  {Tajima}, {Takahashi}, {Thayer}, {Thayer}, {Thompson}, {Tibaldo},
  {Tinivella}, {Torres}, {Tosti}, {Troja}, {Usher}, {Vandenbroucke},
  {Vasileiou}, {Vianello}, {Vitale}, {Winer}, {Wood}, {Yamazaki}, {Younes},
  {Yu}, {Zhu}, {Bhat}, {Briggs}, {Byrne}, {Foley}, {Goldstein}, {Jenke},
  {Kippen}, {Kouveliotou}, {McBreen}, {Meegan}, {Paciesas}, {Preece}, {Rau},
  {Tierney}, {van der Horst}, {von Kienlin}, {Wilson-Hodge}, {Xiong},
  {Cusumano}, {La Parola}, \& {Cummings}}]{ackermann2014}
{Ackermann} M. {et~al.}, 2014, Science, 343, 42

\bibitem[{{Akerlof} {et~al}\mbox{.}(1999){Akerlof}, {Balsano}, {Barthelmy},
  {Bloch}, {Butterworth}, {Casperson}, {Cline}, {Fletcher}, {Frontera},
  {Gisler}, {Heise}, {Hills}, {Kehoe}, {Lee}, {Marshall}, {McKay}, {Miller},
  {Piro}, {Priedhorsky}, {Szymanski}, \& {Wren}}]{akerlof1999}
{Akerlof} C. {et~al.}, 1999, \nat, 398, 400

\bibitem[{{Anderson} {et~al}\mbox{.}(2014){Anderson}, {van der Horst},
  {Staley}, {Fender}, {Wijers}, {Scaife}, {Rumsey}, {Titterington},
  {Rowlinson}, \& {Saunders}}]{anderson2014}
{Anderson} G.~E. {et~al.}, 2014, \mnras, 440, 2059

\bibitem[{{Berger} {et~al}\mbox{.}(2003){Berger}, {Kulkarni}, {Pooley},
  {Frail}, {McIntyre}, {Wark}, {Sari}, {Soderberg}, {Fox}, {Yost}, \&
  {Price}}]{berger2003}
{Berger} E. {et~al.}, 2003, \nat, 426, 154

\bibitem[{{Bernardini} {et~al}\mbox{.}(2014){Bernardini}, {Campana},
  {Ghisellini}, {D'Avanzo}, {Calderone}, {Covino}, {Cusumano}, {Ghirlanda}, {La
  Parola}, {Maselli}, {Melandri}, {Salvaterra}, {Burlon}, {D'Elia}, {Fugazza},
  {Sbarufatti}, {Vergani}, \& {Tagliaferri}}]{bernardini2014}
{Bernardini} M.~G. {et~al.}, 2014, \mnras, 439, L80

\bibitem[{{Bignall} {et~al}\mbox{.}(2006){Bignall}, {Macquart}, {Jauncey},
  {Lovell}, {Tzioumis}, \& {Kedziora-Chudczer}}]{bignall2006}
{Bignall} H.~E., {Macquart} J.-P., {Jauncey} D.~L., {Lovell} J.~E.~J.,
  {Tzioumis} A.~K., {Kedziora-Chudczer} L., 2006, \apj, 652, 1050

\bibitem[{{Cenko} {et~al}\mbox{.}(2011){Cenko}, {Frail}, {Harrison}, {Haislip},
  {Reichart}, {Butler}, {Cobb}, {Cucchiara}, {Berger}, {Bloom}, {Chandra},
  {Fox}, {Perley}, {Prochaska}, {Filippenko}, {Glazebrook}, {Ivarsen},
  {Kasliwal}, {Kulkarni}, {LaCluyze}, {Lopez}, {Morgan}, {Pettini}, \&
  {Rana}}]{cenko2011}
{Cenko} S.~B. {et~al.}, 2011, \apj, 732, 29

\bibitem[{{Chandra} {et~al}\mbox{.}(2008){Chandra}, {Cenko}, {Frail},
  {Chevalier}, {Macquart}, {Kulkarni}, {Bock}, {Bertoldi}, {Kasliwal}, {Fox},
  {Price}, {Berger}, {Soderberg}, {Harrison}, {Gal-Yam}, {Ofek}, {Rau},
  {Schmidt}, {Cameron}, {Cowie}, {Cowie}, {Roth}, {Dopita}, {Peterson}, \&
  {Penprase}}]{chandra2008}
{Chandra} P. {et~al.}, 2008, \apj, 683, 924

\bibitem[{{Chevalier} \& {Li}(2000)}]{chevalier2000}
{Chevalier} R.~A., {Li} Z.-Y., 2000, \apj, 536, 195

\bibitem[{{Cordes} \& {Lazio}(2002)}]{cordes2002}
{Cordes} J.~M., {Lazio} T.~J.~W., 2002, ArXiv Astrophysics e-prints

\bibitem[{{Covino} {et~al}\mbox{.}(1999){Covino}, {Lazzati}, {Ghisellini},
  {Saracco}, {Campana}, {Chincarini}, {di Serego}, {Cimatti}, {Vanzi},
  {Pasquini}, {Haardt}, {Israel}, {Stella}, \& {Vietri}}]{covino1999}
{Covino} S. {et~al.}, 1999, \aap, 348, L1

\bibitem[{{Cucchiara} {et~al}\mbox{.}(2011){Cucchiara}, {Levan}, {Fox},
  {Tanvir}, {Ukwatta}, {Berger}, {Kr{\"u}hler}, {K{\"u}pc{\"u} Yolda{\c s}},
  {Wu}, {Toma}, {Greiner}, {Olivares}, {Rowlinson}, {Amati}, {Sakamoto},
  {Roth}, {Stephens}, {Fritz}, {Fynbo}, {Hjorth}, {Malesani}, {Jakobsson},
  {Wiersema}, {O'Brien}, {Soderberg}, {Foley}, {Fruchter}, {Rhoads},
  {Rutledge}, {Schmidt}, {Dopita}, {Podsiadlowski}, {Willingale}, {Wolf},
  {Kulkarni}, \& {D'Avanzo}}]{cucchiara2011}
{Cucchiara} A. {et~al.}, 2011, \apj, 736, 7

\bibitem[{{Dennett-Thorpe} \& {de Bruyn}(2002)}]{dennettthorpe2002}
{Dennett-Thorpe} J., {de Bruyn} A.~G., 2002, \nat, 415, 57

\bibitem[{{Eichler} {et~al}\mbox{.}(1989){Eichler}, {Livio}, {Piran}, \&
  {Schramm}}]{eichler1989}
{Eichler} D., {Livio} M., {Piran} T., {Schramm} D.~N., 1989, \nat, 340, 126

\bibitem[{{Eichler} \& {Waxman}(2005)}]{eichler2005}
{Eichler} D., {Waxman} E., 2005, \apj, 627, 861

\bibitem[{{Filgas} {et~al}\mbox{.}(2011){Filgas}, {Greiner}, {Schady},
  {Kr{\"u}hler}, {Updike}, {Klose}, {Nardini}, {Kann}, {Rossi}, {Sudilovsky},
  {Afonso}, {Clemens}, {Elliott}, {Nicuesa Guelbenzu}, {Olivares E.}, \&
  {Rau}}]{filgas2011}
{Filgas} R. {et~al.}, 2011, \aap, 535, A57

\bibitem[{{Frail} {et~al}\mbox{.}(2000){Frail}, {Berger}, {Galama}, {Kulkarni},
  {Moriarty-Schieven}, {Pooley}, {Sari}, {Shepherd}, {Taylor}, \&
  {Walter}}]{frail2000a}
{Frail} D.~A. {et~al.}, 2000, \apjl, 538, L129

\bibitem[{{Frail} {et~al}\mbox{.}(1997){Frail}, {Kulkarni}, {Nicastro},
  {Feroci}, \& {Taylor}}]{frail1997}
{Frail} D.~A., {Kulkarni} S.~R., {Nicastro} L., {Feroci} M., {Taylor} G.~B.,
  1997, \nat, 389, 261

\bibitem[{{Frail}, {Waxman} \& {Kulkarni}(2000){Frail}, {Waxman}, \&
  {Kulkarni}}]{frail2000b}
{Frail} D.~A., {Waxman} E., {Kulkarni} S.~R., 2000, \apj, 537, 191

\bibitem[{{Fynbo} {et~al}\mbox{.}(2009){Fynbo}, {Jakobsson}, {Prochaska},
  {Malesani}, {Ledoux}, {de Ugarte Postigo}, {Nardini}, {Vreeswijk},
  {Wiersema}, {Hjorth}, {Sollerman}, {Chen}, {Th{\"o}ne}, {Bj{\"o}rnsson},
  {Bloom}, {Castro-Tirado}, {Christensen}, {De Cia}, {Fruchter}, {Gorosabel},
  {Graham}, {Jaunsen}, {Jensen}, {Kann}, {Kouveliotou}, {Levan}, {Maund},
  {Masetti}, {Milvang-Jensen}, {Palazzi}, {Perley}, {Pian}, {Rol}, {Schady},
  {Starling}, {Tanvir}, {Watson}, {Xu}, {Augusteijn}, {Grundahl}, {Telting}, \&
  {Quirion}}]{fynbo2009}
{Fynbo} J.~P.~U. {et~al.}, 2009, \apjs, 185, 526

\bibitem[{{Fynbo} {et~al}\mbox{.}(2006){Fynbo}, {Watson}, {Th{\"o}ne},
  {Sollerman}, {Bloom}, {Davis}, {Hjorth}, {Jakobsson}, {J{\o}rgensen},
  {Graham}, {Fruchter}, {Bersier}, {Kewley}, {Cassan}, {Castro Cer{\'o}n},
  {Foley}, {Gorosabel}, {Hinse}, {Horne}, {Jensen}, {Klose}, {Kocevski},
  {Marquette}, {Perley}, {Ramirez-Ruiz}, {Stritzinger}, {Vreeswijk}, {Wijers},
  {Woller}, {Xu}, \& {Zub}}]{fynbo2006}
{Fynbo} J.~P.~U. {et~al.}, 2006, \nat, 444, 1047

\bibitem[{{Gehrels} {et~al}\mbox{.}(2006){Gehrels}, {Norris}, {Barthelmy},
  {Granot}, {Kaneko}, {Kouveliotou}, {Markwardt}, {M{\'e}sz{\'a}ros}, {Nakar},
  {Nousek}, {O'Brien}, {Page}, {Palmer}, {Parsons}, {Roming}, {Sakamoto},
  {Sarazin}, {Schady}, {Stamatikos}, \& {Woosley}}]{gehrels2006}
{Gehrels} N. {et~al.}, 2006, \nat, 444, 1044

\bibitem[{{Goodman}(1997)}]{goodman1997}
{Goodman} J., 1997, New Astronomy, 2, 449

\bibitem[{{Granot} \& {K{\"o}nigl}(2003)}]{granot2003}
{Granot} J., {K{\"o}nigl} A., 2003, \apjl, 594, L83

\bibitem[{{Granot} \& {Sari}(2002)}]{granot2002}
{Granot} J., {Sari} R., 2002, \apj, 568, 820

\bibitem[{{Granot} \& {Taylor}(2005)}]{granot2005}
{Granot} J., {Taylor} G.~B., 2005, \apj, 625, 263

\bibitem[{{Granot} \& {van der Horst}(2014)}]{granot2014}
{Granot} J., {van der Horst} A.~J., 2014, \pasa, 31, 8

\bibitem[{{Greiner} {et~al}\mbox{.}(2003){Greiner}, {Klose}, {Reinsch}, {Martin
  Schmid}, {Sari}, {Hartmann}, {Kouveliotou}, {Rau}, {Palazzi}, {Straubmeier},
  {Stecklum}, {Zharikov}, {Tovmassian}, {B{\"a}rnbantner}, {Ries}, {Jehin},
  {Henden}, {Kaas}, {Grav}, {Hjorth}, {Pedersen}, {Wijers}, {Kaufer}, {Park},
  {Williams}, \& {Reimer}}]{greiner2003}
{Greiner} J. {et~al.}, 2003, \nat, 426, 157

\bibitem[{{Itoh} {et~al}\mbox{.}(2013){Itoh}, {Kawaguchi}, {Moritani},
  {Takaki}, {Kawabata}, {Ohno}, {Takahashi}, {Tanaka}, \& {Yoshida}}]{itoh2013}
{Itoh} R. {et~al.}, 2013, GRB Coordinates Network, 14486, 1

\bibitem[{{Jakobsson} {et~al}\mbox{.}(2012){Jakobsson}, {Hjorth}, {Malesani},
  {Chapman}, {Fynbo}, {Tanvir}, {Milvang-Jensen}, {Vreeswijk}, {Letawe}, \&
  {Starling}}]{jakobsson2012}
{Jakobsson} P. {et~al.}, 2012, \apj, 752, 62

\bibitem[{{Kaneko} {et~al}\mbox{.}(2007){Kaneko}, {Ramirez-Ruiz}, {Granot},
  {Kouveliotou}, {Woosley}, {Patel}, {Rol}, {in 't Zand}, {van der Horst},
  {Wijers}, \& {Strom}}]{kaneko2007}
{Kaneko} Y. {et~al.}, 2007, \apj, 654, 385

\bibitem[{{Kobayashi} \& {Sari}(2000)}]{kobayashi2000}
{Kobayashi} S., {Sari} R., 2000, \apj, 542, 819

\bibitem[{{Kouveliotou} {et~al}\mbox{.}(2013){Kouveliotou}, {Granot},
  {Racusin}, {Bellm}, {Vianello}, {Oates}, {Fryer}, {Boggs}, {Christensen},
  {Craig}, {Dermer}, {Gehrels}, {Hailey}, {Harrison}, {Melandri}, {McEnery},
  {Mundell}, {Stern}, {Tagliaferri}, \& {Zhang}}]{kouveliotou2013}
{Kouveliotou} C. {et~al.}, 2013, \apjl, 779, L1

\bibitem[{{Kouveliotou} {et~al}\mbox{.}(1993){Kouveliotou}, {Meegan},
  {Fishman}, {Bhat}, {Briggs}, {Koshut}, {Paciesas}, \&
  {Pendleton}}]{kouveliotou1993}
{Kouveliotou} C., {Meegan} C.~A., {Fishman} G.~J., {Bhat} N.~P., {Briggs}
  M.~S., {Koshut} T.~M., {Paciesas} W.~S., {Pendleton} G.~N., 1993, \apjl, 413,
  L101

\bibitem[{{Kouveliotou}, {Wijers} \& {Woosley}(2012){Kouveliotou}, {Wijers}, \&
  {Woosley}}]{kouveliotou2012}
{Kouveliotou} C., {Wijers} R.~A.~M.~J., {Woosley} S., 2012, {Gamma-ray Bursts}

\bibitem[{{Kulkarni} {et~al}\mbox{.}(1999){Kulkarni}, {Frail}, {Sari},
  {Moriarty-Schieven}, {Shepherd}, {Udomprasert}, {Readhead}, {Bloom},
  {Feroci}, \& {Costa}}]{kulkarni1999}
{Kulkarni} S.~R. {et~al.}, 1999, \apjl, 522, L97

\bibitem[{{Laskar} {et~al}\mbox{.}(2013){Laskar}, {Berger}, {Zauderer},
  {Margutti}, {Soderberg}, {Chakraborti}, {Lunnan}, {Chornock}, {Chandra}, \&
  {Ray}}]{laskar2013}
{Laskar} T. {et~al.}, 2013, \apj, 776, 119

\bibitem[{{Levan} {et~al}\mbox{.}(2013){Levan}, {Tanvir}, {Fruchter}, {Hjorth},
  {Pian}, {Mazzali}, {Perley}, {Cano}, {Graham}, {Hounsell}, {Cenko}, {Fynbo},
  {Kouveliotou}, {Pe'er}, {Misra}, \& {Wiersema}}]{levan2013}
{Levan} A.~J. {et~al.}, 2013, ArXiv e-prints

\bibitem[{{Macquart} \& {de Bruyn}(2007)}]{macquart2007}
{Macquart} J.-P., {de Bruyn} A.~G., 2007, \mnras, 380, L20

\bibitem[{{Maselli} {et~al}\mbox{.}(2014){Maselli}, {Melandri}, {Nava},
  {Mundell}, {Kawai}, {Campana}, {Covino}, {Cummings}, {Cusumano}, {Evans},
  {Ghirlanda}, {Ghisellini}, {Guidorzi}, {Kobayashi}, {Kuin}, {La Parola},
  {Mangano}, {Oates}, {Sakamoto}, {Serino}, {Virgili}, {Zhang}, {Barthelmy},
  {Beardmore}, {Bernardini}, {Bersier}, {Burrows}, {Calderone}, {Capalbi},
  {Chiang}, {D'Avanzo}, {D'Elia}, {De Pasquale}, {Fugazza}, {Gehrels},
  {Gomboc}, {Harrison}, {Hanayama}, {Japelj}, {Kennea}, {Kopac}, {Kouveliotou},
  {Kuroda}, {Levan}, {Malesani}, {Marshall}, {Nousek}, {O'Brien}, {Osborne},
  {Pagani}, {Page}, {Page}, {Perri}, {Pritchard}, {Romano}, {Saito},
  {Sbarufatti}, {Salvaterra}, {Steele}, {Tanvir}, {Vianello}, {Weigand},
  {Wiersema}, {Yatsu}, {Yoshii}, \& {Tagliaferri}}]{maselli2014}
{Maselli} A. {et~al.}, 2014, Science, 343, 48

\bibitem[{{Matsumiya} \& {Ioka}(2003)}]{matsumiya2003}
{Matsumiya} M., {Ioka} K., 2003, \apjl, 595, L25

\bibitem[{{M{\'e}sz{\'a}ros} \& {Rees}(1999)}]{meszaros1999}
{M{\'e}sz{\'a}ros} P., {Rees} M.~J., 1999, \mnras, 306, L39

\bibitem[{{Mundell} {et~al}\mbox{.}(2013){Mundell}, {Kopa{\v c}}, {Arnold},
  {Steele}, {Gomboc}, {Kobayashi}, {Harrison}, {Smith}, {Guidorzi}, {Virgili},
  {Melandri}, \& {Japelj}}]{mundell2013}
{Mundell} C.~G. {et~al.}, 2013, \nat, 504, 119

\bibitem[{{Narayan}, {Paczynski} \& {Piran}(1992){Narayan}, {Paczynski}, \&
  {Piran}}]{narayan1992}
{Narayan} R., {Paczynski} B., {Piran} T., 1992, \apjl, 395, L83

\bibitem[{{Panaitescu}, {Vestrand} \& {Wo{\'z}niak}(2013){Panaitescu},
  {Vestrand}, \& {Wo{\'z}niak}}]{panaitescu2013}
{Panaitescu} A., {Vestrand} W.~T., {Wo{\'z}niak} P., 2013, \mnras, 436, 3106

\bibitem[{{Pedersen} {et~al}\mbox{.}(1998){Pedersen}, {Jaunsen}, {Grav},
  {Ostensen}, {Andersen}, {Wold}, {Kristen}, {Broeils}, {Naeslund}, {Fransson},
  {Lacy}, {Castro-Tirado}, {Gorosabel}, {Rodriguez Espinosa}, {Perez}, {Wolf},
  {Fockenbrock}, {Hjorth}, {Muhli}, {Hakala}, {Piro}, {Feroci}, {Costa},
  {Nicastro}, {Palazzi}, {Frontera}, {Monaldi}, \& {Heise}}]{pedersen1998}
{Pedersen} H. {et~al.}, 1998, \apj, 496, 311

\bibitem[{{Peng}, {K{\"o}nigl} \& {Granot}(2005){Peng}, {K{\"o}nigl}, \&
  {Granot}}]{peng2005}
{Peng} F., {K{\"o}nigl} A., {Granot} J., 2005, \apj, 626, 966

\bibitem[{{Perley} {et~al}\mbox{.}(2014){Perley}, {Cenko}, {Corsi}, {Tanvir},
  {Levan}, {Kann}, {Sonbas}, {Wiersema}, {Zheng}, {Zhao}, {Bai}, {Bremer},
  {Castro-Tirado}, {Chang}, {Clubb}, {Frail}, {Fruchter}, {G{\"o}{\u g}{\"u}{\c
  s}}, {Greiner}, {G{\"u}ver}, {Horesh}, {Filippenko}, {Klose}, {Mao},
  {Morgan}, {Pozanenko}, {Schmidl}, {Stecklum}, {Tanga}, {Volnova}, {Volvach},
  {Wang}, {Winters}, \& {Xin}}]{perley2014}
{Perley} D.~A. {et~al.}, 2014, \apj, 781, 37

\bibitem[{{Preece} {et~al}\mbox{.}(2014){Preece}, {Burgess}, {von Kienlin},
  {Bhat}, {Briggs}, {Byrne}, {Chaplin}, {Cleveland}, {Collazzi}, {Connaughton},
  {Diekmann}, {Fitzpatrick}, {Foley}, {Gibby}, {Giles}, {Goldstein}, {Greiner},
  {Gruber}, {Jenke}, {Kippen}, {Kouveliotou}, {McBreen}, {Meegan}, {Paciesas},
  {Pelassa}, {Tierney}, {van der Horst}, {Wilson-Hodge}, {Xiong}, {Younes},
  {Yu}, {Ackermann}, {Ajello}, {Axelsson}, {Baldini}, {Barbiellini}, {Baring},
  {Bastieri}, {Bellazzini}, {Bissaldi}, {Bonamente}, {Bregeon}, {Brigida},
  {Bruel}, {Buehler}, {Buson}, {Caliandro}, {Cameron}, {Caraveo}, {Cecchi},
  {Charles}, {Chekhtman}, {Chiang}, {Chiaro}, {Ciprini}, {Claus},
  {Cohen-Tanugi}, {Cominsky}, {Conrad}, {D'Ammando}, {de Angelis}, {de Palma},
  {Dermer}, {Desiante}, {Digel}, {Di Venere}, {Drell}, {Drlica-Wagner},
  {Favuzzi}, {Franckowiak}, {Fukazawa}, {Fusco}, {Gargano}, {Gehrels},
  {Germani}, {Giglietto}, {Giordano}, {Giroletti}, {Godfrey}, {Granot},
  {Grenier}, {Guiriec}, {Hadasch}, {Hanabata}, {Harding}, {Hayashida},
  {Iyyani}, {Jogler}, {J{\'o}hannesson}, {Kawano}, {Kn{\"o}dlseder},
  {Kocevski}, {Kuss}, {Lande}, {Larsson}, {Larsson}, {Latronico}, {Longo},
  {Loparco}, {Lovellette}, {Lubrano}, {Mayer}, {Mazziotta}, {Michelson},
  {Mizuno}, {Monzani}, {Moretti}, {Morselli}, {Murgia}, {Nemmen}, {Nuss},
  {Nymark}, {Ohno}, {Ohsugi}, {Okumura}, {Omodei}, {Orienti}, {Paneque},
  {Perkins}, {Pesce-Rollins}, {Piron}, {Pivato}, {Porter}, {Racusin},
  {Rain{\`o}}, {Rando}, {Razzano}, {Razzaque}, {Reimer}, {Reimer}, {Ritz},
  {Roth}, {Ryde}, {Sartori}, {Scargle}, {Schulz}, {Sgr{\`o}}, {Siskind},
  {Spandre}, {Spinelli}, {Suson}, {Tajima}, {Takahashi}, {Thayer}, {Thayer},
  {Tibaldo}, {Tinivella}, {Torres}, {Tosti}, {Troja}, {Usher}, {Vandenbroucke},
  {Vasileiou}, {Vianello}, {Vitale}, {Werner}, {Winer}, {Wood}, \&
  {Zhu}}]{preece2014}
{Preece} R. {et~al.}, 2014, Science, 343, 51

\bibitem[{{Racusin} {et~al}\mbox{.}(2008){Racusin}, {Karpov}, {Sokolowski},
  {Granot}, {Wu}, {Pal'Shin}, {Covino}, {van der Horst}, {Oates}, {Schady},
  {Smith}, {Cummings}, {Starling}, {Piotrowski}, {Zhang}, {Evans}, {Holland},
  {Malek}, {Page}, {Vetere}, {Margutti}, {Guidorzi}, {Kamble}, {Curran},
  {Beardmore}, {Kouveliotou}, {Mankiewicz}, {Melandri}, {O'Brien}, {Page},
  {Piran}, {Tanvir}, {Wrochna}, {Aptekar}, {Barthelmy}, {Bartolini}, {Beskin},
  {Bondar}, {Bremer}, {Campana}, {Castro-Tirado}, {Cucchiara}, {Cwiok},
  {D'Avanzo}, {D'Elia}, {Della Valle}, {de Ugarte Postigo}, {Dominik},
  {Falcone}, {Fiore}, {Fox}, {Frederiks}, {Fruchter}, {Fugazza}, {Garrett},
  {Gehrels}, {Golenetskii}, {Gomboc}, {Gorosabel}, {Greco}, {Guarnieri},
  {Immler}, {Jelinek}, {Kasprowicz}, {La Parola}, {Levan}, {Mangano}, {Mazets},
  {Molinari}, {Moretti}, {Nawrocki}, {Oleynik}, {Osborne}, {Pagani}, {Pandey},
  {Paragi}, {Perri}, {Piccioni}, {Ramirez-Ruiz}, {Roming}, {Steele}, {Strom},
  {Testa}, {Tosti}, {Ulanov}, {Wiersema}, {Wijers}, {Winters}, {Zarnecki},
  {Zerbi}, {M{\'e}sz{\'a}ros}, {Chincarini}, \& {Burrows}}]{racusin2008}
{Racusin} J.~L. {et~al.}, 2008, \nat, 455, 183

\bibitem[{{Ramirez-Ruiz}, {Celotti} \& {Rees}(2002){Ramirez-Ruiz}, {Celotti},
  \& {Rees}}]{ramirezruiz2002}
{Ramirez-Ruiz} E., {Celotti} A., {Rees} M.~J., 2002, \mnras, 337, 1349

\bibitem[{{Rickett}(1990)}]{rickett1990}
{Rickett} B.~J., 1990, \araa, 28, 561

\bibitem[{{Sari} \& {Piran}(1995)}]{sari1995}
{Sari} R., {Piran} T., 1995, \apjl, 455, L143

\bibitem[{{Sari}, {Piran} \& {Narayan}(1998){Sari}, {Piran}, \&
  {Narayan}}]{sari1998}
{Sari} R., {Piran} T., {Narayan} R., 1998, \apjl, 497, L17

\bibitem[{{Sault}, {Teuben} \& {Wright}(1995){Sault}, {Teuben}, \&
  {Wright}}]{sault1995}
{Sault} R.~J., {Teuben} P.~J., {Wright} M.~C.~H., 1995, in Astronomical Society
  of the Pacific Conference Series, Vol.~77, Astronomical Data Analysis
  Software and Systems IV, {Shaw} R.~A., {Payne} H.~E., {Hayes} J.~J.~E., eds.,
  p. 433

\bibitem[{{Shepherd}, {Pearson} \& {Taylor}(1994){Shepherd}, {Pearson}, \&
  {Taylor}}]{shepherd1994}
{Shepherd} M.~C., {Pearson} T.~J., {Taylor} G.~B., 1994, in Bulletin of the
  American Astronomical Society, Vol.~26, Bulletin of the American Astronomical
  Society, pp. 987--989

\bibitem[{{Starling} {et~al}\mbox{.}(2011){Starling}, {Wiersema}, {Levan},
  {Sakamoto}, {Bersier}, {Goldoni}, {Oates}, {Rowlinson}, {Campana},
  {Sollerman}, {Tanvir}, {Malesani}, {Fynbo}, {Covino}, {D'Avanzo}, {O'Brien},
  {Page}, {Osborne}, {Vergani}, {Barthelmy}, {Burrows}, {Cano}, {Curran}, {de
  Pasquale}, {D'Elia}, {Evans}, {Flores}, {Fruchter}, {Garnavich}, {Gehrels},
  {Gorosabel}, {Hjorth}, {Holland}, {van der Horst}, {Hurkett}, {Jakobsson},
  {Kamble}, {Kouveliotou}, {Kuin}, {Kaper}, {Mazzali}, {Nugent}, {Pian},
  {Stamatikos}, {Th{\"o}ne}, \& {Woosley}}]{starling2011}
{Starling} R.~L.~C. {et~al.}, 2011, \mnras, 411, 2792

\bibitem[{{Starling} {et~al}\mbox{.}(2005){Starling}, {Wijers}, {Hughes},
  {Tanvir}, {Vreeswijk}, {Rol}, \& {Salamanca}}]{starling2005}
{Starling} R.~L.~C., {Wijers} R.~A.~M.~J., {Hughes} M.~A., {Tanvir} N.~R.,
  {Vreeswijk} P.~M., {Rol} E., {Salamanca} I., 2005, \mnras, 360, 305

\bibitem[{{Steele} {et~al}\mbox{.}(2009){Steele}, {Mundell}, {Smith},
  {Kobayashi}, \& {Guidorzi}}]{steele2009}
{Steele} I.~A., {Mundell} C.~G., {Smith} R.~J., {Kobayashi} S., {Guidorzi} C.,
  2009, \nat, 462, 767

\bibitem[{{Tan}(1991)}]{tan1991}
{Tan} G.~H., 1991, in Astronomical Society of the Pacific Conference Series,
  Vol.~19, IAU Colloq. 131: Radio Interferometry. Theory, Techniques, and
  Applications, {Cornwell} T.~J., {Perley} R.~A., eds., pp. 42--46

\bibitem[{{Taylor} {et~al}\mbox{.}(2004){Taylor}, {Frail}, {Berger}, \&
  {Kulkarni}}]{taylor2004}
{Taylor} G.~B., {Frail} D.~A., {Berger} E., {Kulkarni} S.~R., 2004, \apjl, 609,
  L1

\bibitem[{{Taylor} {et~al}\mbox{.}(2005){Taylor}, {Momjian}, {Pihlstr{\"o}m},
  {Ghosh}, \& {Salter}}]{taylor2005}
{Taylor} G.~B., {Momjian} E., {Pihlstr{\"o}m} Y., {Ghosh} T., {Salter} C.,
  2005, \apj, 622, 986

\bibitem[{{Toma}, {Ioka} \& {Nakamura}(2008){Toma}, {Ioka}, \&
  {Nakamura}}]{toma2008}
{Toma} K., {Ioka} K., {Nakamura} T., 2008, \apjl, 673, L123

\bibitem[{{van der Horst}(2007)}]{vanderhorst2007}
{van der Horst} A.~J., 2007, PhD thesis, University of Amsterdam

\bibitem[{{van Moorsel}, {Kemball} \& {Greisen}(1996){van Moorsel}, {Kemball},
  \& {Greisen}}]{vanmoorsel1996}
{van Moorsel} G., {Kemball} A., {Greisen} E., 1996, in Astronomical Society of
  the Pacific Conference Series, Vol. 101, Astronomical Data Analysis Software
  and Systems V, {Jacoby} G.~H., {Barnes} J., eds., p.~37

\bibitem[{{Vestrand} {et~al}\mbox{.}(2014){Vestrand}, {Wren}, {Panaitescu},
  {Wozniak}, {Davis}, {Palmer}, {Vianello}, {Omodei}, {Xiong}, {Briggs},
  {Elphick}, {Paciesas}, \& {Rosing}}]{vestrand2014}
{Vestrand} W.~T. {et~al.}, 2014, Science, 343, 38

\bibitem[{{Vink} \& {de Koter}(2005)}]{vink2005}
{Vink} J.~S., {de Koter} A., 2005, \aap, 442, 587

\bibitem[{{Vink}, {de Koter} \& {Lamers}(2001){Vink}, {de Koter}, \&
  {Lamers}}]{vink2001}
{Vink} J.~S., {de Koter} A., {Lamers} H.~J.~G.~L.~M., 2001, \aap, 369, 574

\bibitem[{{Vlahakis}, {Peng} \& {K{\"o}nigl}(2003){Vlahakis}, {Peng}, \&
  {K{\"o}nigl}}]{vlahakis2003}
{Vlahakis} N., {Peng} F., {K{\"o}nigl} A., 2003, \apjl, 594, L23

\bibitem[{{Walker}(1998)}]{walker1998}
{Walker} M.~A., 1998, \mnras, 294, 307

\bibitem[{{Wiersema} {et~al}\mbox{.}(2014){Wiersema}, {Covino}, {Toma}, {van
  der Horst}, {Varela}, {Min}, {Greiner}, {Starling}, {Tanvir}, {Wijers},
  {Campana}, {Curran}, {Fan}, {Fynbo}, {Gorosabel}, {Gomboc}, {Gotz}, {Hjorth},
  {Jin}, {Kobayashi}, {Kouveliotou}, {Mundell}, {O'Brien}, {Pian}, {Rowlinson},
  {Russell}, {Salvaterra}, {di Serego Alighieri}, {Tagliaferri}, {Vergani},
  {Elliott}, {Farina}, {Hartoog}, {Karjalainen}, {Klose}, {Knust}, {Levan},
  {Schady}, {Sudilovsky}, \& {Willingale}}]{wiersema2014}
{Wiersema} K. {et~al.}, 2014, \nat, 509, 201

\bibitem[{{Wiersema} {et~al}\mbox{.}(2012){Wiersema}, {Curran}, {Kr{\"u}hler},
  {Melandri}, {Rol}, {Starling}, {Tanvir}, {van der Horst}, {Covino}, {Fynbo},
  {Goldoni}, {Gorosabel}, {Hjorth}, {Klose}, {Mundell}, {O'Brien}, {Palazzi},
  {Wijers}, {D'Elia}, {Evans}, {Filgas}, {Gomboc}, {Greiner}, {Guidorzi},
  {Kaper}, {Kobayashi}, {Kouveliotou}, {Levan}, {Rossi}, {Rowlinson}, {Steele},
  {de Ugarte Postigo}, \& {Vergani}}]{wiersema2012}
{Wiersema} K. {et~al.}, 2012, \mnras, 426, 2

\bibitem[{{Wijers} {et~al}\mbox{.}(1999){Wijers}, {Vreeswijk}, {Galama}, {Rol},
  {van Paradijs}, {Kouveliotou}, {Giblin}, {Masetti}, {Palazzi}, {Pian},
  {Frontera}, {Nicastro}, {Falomo}, {Soffitta}, \& {Piro}}]{wijers1999}
{Wijers} R.~A.~M.~J. {et~al.}, 1999, \apjl, 523, L33

\bibitem[{{Woosley}(1993)}]{woosley1993}
{Woosley} S.~E., 1993, \apj, 405, 273

\bibitem[{{Woosley} \& {Heger}(2006)}]{woosley2006}
{Woosley} S.~E., {Heger} A., 2006, \apj, 637, 914

\bibitem[{{Xu} {et~al}\mbox{.}(2013){Xu}, {de Ugarte Postigo}, {Leloudas},
  {Kr{\"u}hler}, {Cano}, {Hjorth}, {Malesani}, {Fynbo}, {Th{\"o}ne},
  {S{\'a}nchez-Ram{\'{\i}}rez}, {Schulze}, {Jakobsson}, {Kaper}, {Sollerman},
  {Watson}, {Cabrera-Lavers}, {Cao}, {Covino}, {Flores}, {Geier}, {Gorosabel},
  {Hu}, {Milvang-Jensen}, {Sparre}, {Xin}, {Zhang}, {Zheng}, \& {Zou}}]{xu2013}
{Xu} D. {et~al.}, 2013, \apj, 776, 98

\bibitem[{{Yi}, {Wu} \& {Dai}(2013){Yi}, {Wu}, \& {Dai}}]{yi2013}
{Yi} S.-X., {Wu} X.-F., {Dai} Z.-G., 2013, \apj, 776, 120

\bibitem[{{Zou}, {Wu} \& {Dai}(2005){Zou}, {Wu}, \& {Dai}}]{zou2005}
{Zou} Y.~C., {Wu} X.~F., {Dai} Z.~G., 2005, \mnras, 363, 93

\end{thebibliography}

\end{document}